\RequirePackage{fix-cm}
\documentclass[twocolumn]{svjour3}          
\smartqed  
\usepackage{graphicx}
\usepackage{natbib}
\usepackage{amsmath}
\usepackage{upgreek}
\usepackage{hyperref}
\hypersetup{
             colorlinks=true,
             linkcolor=black,
             filecolor=black,
             citecolor=black,
             urlcolor=blue
           }
\newcommand{\Frac}[2]{\frac{\displaystyle\strut #1}{\displaystyle\strut #2} }
\renewcommand{\tau}{\uptau}

\usepackage{ulem}
\usepackage{xcolor}
\definecolor{Dred}{rgb}{0.312,0.070,0.070}
\definecolor{redc}{rgb}{0.999,0.000,0.000}

\newcommand{\Red}[1]{\textcolor{redc}{\textbf{#1}}}
\newcounter{nnote}

\let\oldmarginpar\marginpar
\renewcommand\marginpar[1]{\-\oldmarginpar[\raggedleft\footnotesize #1]%
{\raggedright\footnotesize #1}}

\newcommand\strike{\bgroup\markoverwith{\textcolor{red}{\rule[0.5ex]{2pt}{1.0pt}}}\ULon}
\newcommand{\web}[1]{\url{#1}}
\newcommand{\Strike}[1]{\Red{#1}}

\renewcommand{\strike}[1]{\relax}
\renewcommand{\Strike}[1]{\relax}
\journalname{Journal of Geodesy}
\begin{document}

\title{The use of astronomy VLBA campaign MOJAVE for geodesy
}



\author{Hana Kr\'asn\'a         \and
        Leonid Petrov 
}


\institute{H. Kr\'asn\'a \at
  1. Department of Geodesy and Geoinformation, Technische Universit\"at Wien, Vienna, Austria \\
  2. Astronomical Institute of the Czech Academy of Sciences, Prague, Czech Republic\\
  \email{hana.krasna@tuwien.ac.at}            \\
  \and
  L. Petrov \at
  NASA Goddard Space Flight Center, Code 61A, Greenbelt, USA
}

\date{Received: 02 May 2021 / Accepted: date}

\maketitle

\begin{abstract}
We investigated the suitability of the astronomical 15 GHz
VLBA observing program MOJAVE-5 for estimation of geodetic
parameters, such as station coordinates and Earth orientation
parameters. We processed contemporary geodetic dual-band RV and
CN experiments observed at 2.3 GHz and 8.6 GHz starting on
September 2016 through July 2020 as reference dataset. We showed
that the baseline length repeatability from MOJAVE-5 experiments
is only a factor of 1.5 greater than from the dedicated geodetic
dataset and still below 1~ppb. The wrms of the difference of
estimated EOP with respect to the reference IERS C04 time series
are a factor of 1.3 to 1.8 worse. We isolated three major
differences between the datasets in terms their possible impact
on the geodetic results, i.e. the scheduling approach, treatment
of the ionospheric delay, and selection of target radio sources.
We showed that the major factor causing discrepancies in the
estimated geodetic parameters is the different scheduling
approach of the datasets. We conclude that systematic errors
in MOJAVE-5 dataset are low enough for these data to be used
as an excellent testbed for further investigations on the
radio source structure effects in geodesy and astrometry.

\end{abstract}

\section{Introduction}
\label{intro}

  Group delay of an extended source observed with very long
baseline interferometry (VLBI) differs from the group delay
of a point source. Up to now, the contribution of source
structure is not included in routine analysis of VLBI data.
It was known for long time that source structure is
a significant \citep[e.g.,][]{r:zep93,Sovers02, Tornatore07,
Shabala15, r:gaia3} or even the major~\citep{Anderson18} contributor
to the error budget in geodetic VLBI.

  One of the most promising ways to compute the source
structure contribution to group delay is to generate
images from the same VLBI observations, perform their 2D
Fourier transform over spatial coordinates, and use it for
calculation of structure delay \citep[see, e.g.,][]{r:gaia3}. Unfortunately,
geodetic observing schedules are not well suited for producing
good quality images. A typical geodetic schedule splits the network
into a number of ad hoc subarrays, so a subset of stations observes
one source and a subset of other stations observes another source
at the same time, and upon completion of integration another
subset of stations observes the next source. This leads to
a substantial reduction of the number of closures in phase
and amplitude required for robust imaging. Astronomical schedules
usually avoid subarrays. The use of data for geodesy and
astrometry from astronomical programs designed for imaging was
not common in the past because four to eight intermediate recorded
frequencies (IFs) were usually allocated contiguously, while for
geodetic applications the frequencies are allocated as wide as
possible. As a result, group delay uncertainty at a given signal
to noise ratio (SNR) was an order of magnitude worse than from
geodetic schedules. Although such data were still useful
for astrometry \citep{r:obrs1,r:obrs2}, they were too coarse
for precise geodesy. A non-contiguous allocation of intermediate
frequencies for astronomy projects was rare because usually it was
not required, and a commonly used AIPS software package \citep{r:aips}
that implemented the fringe fitting procedure does not support direct
processing of such data. In a case when a goal of astronomical
observation requires wide spanned bandwidth, e.g., for VLBA (Very
Long Baseline Array) Imaging and Polarimetry Survey at 5~GHz
\citep{r:vips}, processing astronomical data in a geodetic/astrometric
mode was feasible and provided good results \citep{r:astro_vips}.
However, single-band  observations at rather low frequencies such
as 5~GHz are affected by the ionospheric contribution, and this
limits their usability for geodesy.

  Progress in radioastronomy instrumentation resulted in
an increase of recorded bandwidth. Since 2016--2020 astronomical
observations typically cover a frequency band of 256 or 512~MHz.
Group delay precision from these setups is close to the precision
reached at geodetic setups. Therefore, the use of astronomical
observing program seems feasible as a testbed for studying source
structure contribution in detail provided that such a program satisfies
two other remaining criteria: a) it observes strong sources and b) it is
conducted at rather high frequencies to minimize the impact of the
ionosphere. MOJAVE-5 (Monitoring Of Jets in Active galactic nuclei
with VLBA Experiments) suits both these criteria. The program started
in 1994 \citep{Lister18} and is focused on observations of bright
active galactic nuclea (AGNs) with discernible structure at 15~GHz.

\section{Motivation}
\label{motiv}

  Before commencing a thorough investigation of the impact of source
structure on astrometry and geodesy results, we need establish a solid
foundation of that work. MOJAVE-5 dataset differs from an usual geodetic
dataset a)~by the way how it was scheduled; b)~by observing frequencies;
and c)~by the source selection.

  An observing schedule consists of a sequence of time intervals
called scans when all or a part of antennas of the network record voltage
from a given source. Astronomical schedules are usually made by
optimization of the $uv$-coverage, i.e. projections of the baseline
vector on the plane tangential to the source direction. The scheduling
goal of astronomical experiments is to generate such a sequence of
observations that covers that plane as uniform as possible for each
program source. Geodetic schedules are usually designed to optimize
elevation/azimuth coverage at each station for short time intervals
(1--3 hours).

  Geodetic observations are done at two or more frequencies
simultaneously. Since the ionospheric group delay is frequency
dependent, multi-band observations allow to derive an ionosphere
free combination of group delays. Astronomical observations are usually
done at one frequency at once. Therefore, group delay observables
from astronomical observations are affected by the ionosphere.

  A list of $\sim\!\!$ 100 objects is usually observed with
geodesy schedules. Sources with extended structures are observed
less often than point-like sources. Astronomical schedules have
less sources, but they are observed more intensively during an
experiment and sources with extended structures are preferably picked.

  We want to answer the following questions in this paper:
1)~what are the metrics of geodetic parameters derived from the MOJAVE-5
dataset? 2)~how worse or better these metrics are with respect to similar
geodetic programs? 3)~what is the main cause of these differences?
And finally, we want to learn whether we can use MOJAVE-5 dataset
as a testbed for investigation of the impact of source structure
on geodetic and astrometric results.

\section{Data analysis}
\label{sec:da}

\begin{figure}
  \includegraphics[trim=6cm 0cm 0cm 0cm, clip=true, angle=270, width=0.5\textwidth]{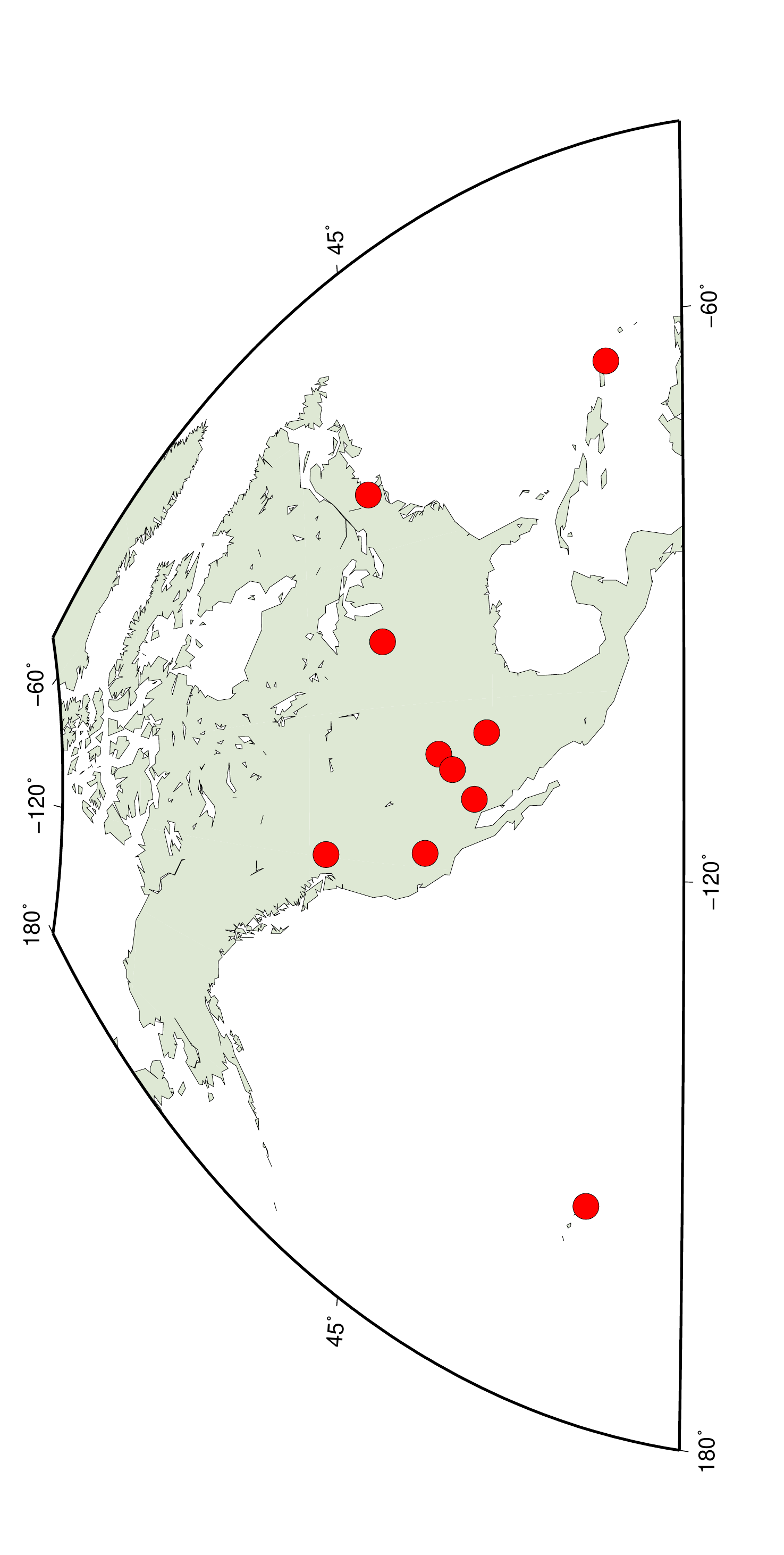}
  \caption{Distribution of the ten VLBA radio telescopes.}
  \label{fig:vlba}       
\end{figure}

The VLBA network consists of ten 25-meters radio telescopes located
on the U.S. territory (eight in North America, one in the Pacific,
and one in the Caribbean), see Fig.~\ref{fig:vlba}. The interferometric
visibility data of MOJAVE-5 campaign (observing code bl229) at
15.3~GHz (Ku band) with dual circular polarization are publicly available through
the National Radio Astronomy Observatory (NRAO) Science Data
archive\footnote{\href{https://archive.nrao.edu/archive}{https://archive.nrao.edu/archive}}
in the FITS-IDI (Interferometry Data Interchange) format. We processed 33 MOJAVE-5 experiments since September 26, 2016 through July 02, 2020. The first 25~experiments (bl229aa-ay) were observed
at independently recorded eight intermediate frequencies of 32~MHz width per polarization using
the polyphase filter bank (PFB) personality of the digital backend. Since July 2019 (experiment bl229az) the bandwidth of a sub-band has increased to 64~MHz, which
built four sub-bands covering 128~channels and MOJAVE-5 campaign has used four
IFs of 64~MHz width per polarization using the direct digital converter (DDC) personality of the digital
backend. In both cases the total recorded bandwidth per polarization is
256~MHz (see Table~\ref{tab:bl229channels}).

\begin{table}
\caption{Lower edge frequency of the sub-bands in the MOJAVE-5 bl229 experiments in GHz.}
\label{tab:bl229channels}       
\begin{tabular}{ll}
\hline\noalign{\smallskip}
bl229aa--ay & bl229az--bg  \\
\noalign{\smallskip}\hline\noalign{\smallskip}
15.22400 & 15.17575\\
15.25600 & 15.25575\\
15.28800 & 15.31975\\
15.32000 & 15.38375\\
15.35200 & \\
15.38400 & \\
15.41600 & \\
15.44800 & \\
\noalign{\smallskip}\hline
\end{tabular}
\end{table}

We processed the observations with the fringe-fitting software PIMA \citep{Petrov11}
running coarse fringe fitting -- bandpass calibration -- fine fringe fitting
and producing the output databases that include group delays and their
uncertainties among other parameters. These quantities  serve as input for
the data analysis software package
pSolve\footnote{\href{http://astrogeo.org/psolve}{http://astrogeo.org/psolve}}
and VieVS \citep{Boehm18}.

  As a reference dataset, we analyzed 28~geodetic RV (Regular geodesy with VLBI)
\citep{r:rdv} and 6~CN sessions observed simultaneously at 2.3/8.6~GHz (S/X bands)
for the same time span starting with rv119 on September 14, 2016 through July 07,
2020. The RV network consists of ten VLBA stations plus up to seven other geodetic
stations. These sessions are designed to provide accurate estimates of the Earth
orientation parameters (EOP), a highly accurate terrestrial reference frame
(TRF) determination, and source position estimation where the VLBA stations are
incorporated into the VLBI reference frame through the inclusion of other geodetic stations
with long history of observations. In CN experiments only ten VLBA stations
participate \citep{thomas20}.

  To assess the quality of geodetic results, we estimated baseline lengths and EOP
and computed their weighted root mean squares (wrms) from the astrophysics MOJAVE-5
program and dedicated geodetic RV\&CN experiments. As an extra check, we
analyzed the VLBA data in several ways. One solution was produced using the software
PIMA for the fringe-fitting and pSolve for the analysis. In the second solution we
analyzed group delays produced with PIMA with the geodetic VLBI analysis software
VieVS. For the RV\&CN experiments we run another solution with VieVS
where we used group delays evaluated with Fourfit visibility analysis
software \citep{Cappallo17}. These data products in vgosDB format were retrieved from the
International VLBI Service for Geodesy \& Astrometry (IVS) data
archive\footnote{Available at \href{https://ivscc.gsfc.nasa.gov/products-data}
{https://ivscc.gsfc.nasa.gov/products-data}}.
Tables~\ref{tab:est_param_psolve} and~\ref{tab:est_param_vievs} contain the
parameterization of the solutions in pSolve and VieVS, respectively.
Table~\ref{tab:wrms_ses} shows weighted rms of postfit residuals.
The MOJAVE-5 and RV\&CN experiments are processed in the same manner with
the same parameterization to allow an informative comparison.

\begin{table*}
    \caption{Parameterization of estimated parameters of a single session solutions in pSolve.}
    \label{tab:est_param_psolve}
    \begin{tabular}{ll}
       \hline\noalign{\smallskip}
       \multicolumn{2}{c}{pSolve}  \\
    \noalign{\smallskip}\hline\noalign{\smallskip}
      CRF  & selected sources \\
      TRF  & NNT/NNR condition on VLBA stations with 0.1 mm constraints \\
      ERP  & offset and rate \\
             celestial pole offsets & offset without constraints\\
      zenith wet delay & B-spline with time span 20 min and sigma of constraints 50 ps/h\\
      tropo. gradients &8 hours with sigma of constr. 0.5 mm on offset and 2.00 mm/day on rate\\
      clocks & B-spline with time span 60 min and constraint sigma 5.e-14 s/s\\
      baseline clock offsets  & offset with constraint sigma 500 ns \\
      weights & yes \\
      \noalign{\smallskip}\hline
     \end{tabular}
\end{table*}

\begin{table*}
\caption{Parameterization of estimated parameters of a single session solutions in VieVS.}
\label{tab:est_param_vievs}       
\begin{tabular}{ll}
\hline\noalign{\smallskip}
   \multicolumn{2}{c}{VieVS}  \\
\noalign{\smallskip}\hline\noalign{\smallskip}
CRF & selected sources \\
TRF & NNT/NNR condition on VLBA stations  \\
ERP  & pwlo with time interval 24 hours with relative constraints 1~mas \\
celestial pole offsets & pwlo with time interval 24 hours with relative constraints 0.1~$\mu$as \\
zenith wet delay & pwlo with time interval 30 min with relative constraints 1.5~cm\\
tropo. gradients &pwlo with time interval 180 min with relative constraints 0.5~cm\\
clocks &pwlo with time interval 60 min with relative constraints 1.3 cm, one rate and
	   quadratic term \\
baseline clock offsets  & offset without constraints \\
weights & baseline-dependent weighting \\
\noalign{\smallskip}\hline
\end{tabular}
\end{table*}

\begin{table}
     \caption{Weighted rms of post fit residuals in ps.}
     \label{tab:wrms_ses}
     \begin{tabular}{lllc}
          \hline\noalign{\smallskip}
          & min & max & median \\
          \noalign{\smallskip}\hline\noalign{\smallskip}
          MOJAVE-5 bl229 series     & 11.3 & 28.7 & 18.4 \\
          geodetic RV\&CN all stat  & 14.7 & 40.3 & 25.2 \\
          geodetic RV\&CN VLBA only & 14.7 & 37.8 & 24.1 \\
          \noalign{\smallskip}\hline
     \end{tabular}
\end{table}

\paragraph{Baseline length repeatability.}

\begin{figure*}
  \includegraphics[width=0.49\textwidth]{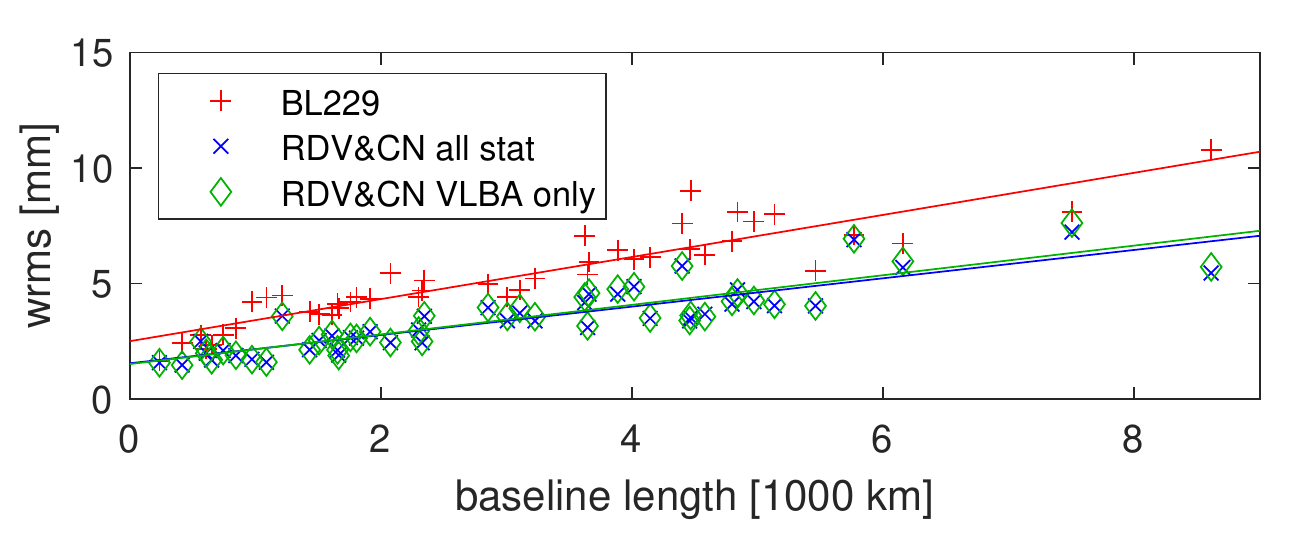}
  \includegraphics[width=0.49\textwidth]{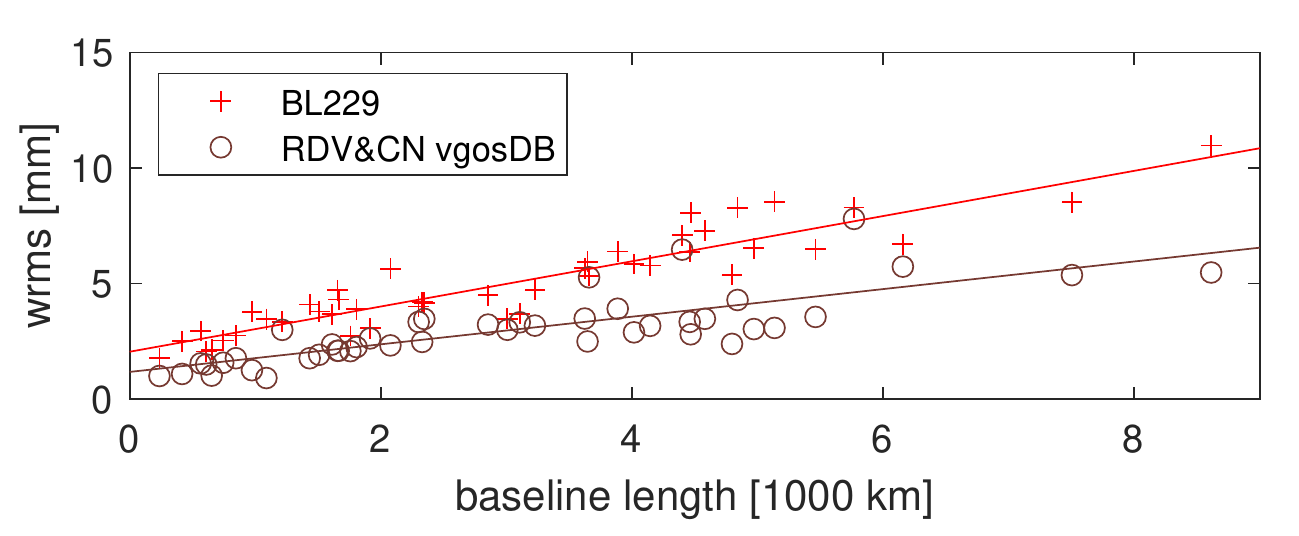}
  \caption{Baseline length repeatability at the VLBA network. Left panel
           compares baseline scatter computed with pSolve from MOJAVE-5 dataset
           (red crosses), RV\&CN dataset after processing data from all stations
           (blue x-signs), and RV\&CN dataset when non-VLBA observations were
           dropped (green-diamonds). Right panel shows baseline length
           repeatability computed with VieVS from MOJAVE-5 dataset (red
           crosses) and RV\&CN dataset processed with Fourfit (brown circles).
           }
\label{fig:blrL2}
\end{figure*}

We ran several solutions and we compared the scatter in baseline length estimates.
We show in Fig.~\ref{fig:blrL2} the wrms of the estimated baseline lengths
from solutions computed with pSolve (left panel) and with VieVS
(right panel). Red crosses denote the baselines determined from the MOJAVE-5
experiments in both plots. We show in the left plot of Fig.~\ref{fig:blrL2}
the baseline scatter computed from the RV\&CN sessions with the whole
scheduled network (blue x-signs) and with observations conducted at
the VLBA stations only (green diamonds). The plot demonstrates that dropping
the data obtained with non-VLBA stations does not change the wrms of the baseline
lengths between the VLBA telescopes. In the right plot of that figure we compare
the MOJAVE-5 bl229 baseline scatter with RV\&CN sessions processed with Fourfit.
We got approximately the same baseline length scatter from MOJAVE-5 and RV\&CN
sessions using independent software packages.

  There is an increase in the baseline length repeatability from
a solution using the MOJAVE-5 dataset with respect to the reference
RV\&CN sessions. The baseline length repeatability differences
derived from RV\&CN and MOJAVE-5 are about 1.3~mm at a 1000~km long
baseline and 3.2~mm at the 8611~km baseline length. The coefficients
of the linear regression are summarized in Table~\ref{tab:bsl}.
We conclude the baseline length repeatability derived from analysis of
single-band 15~GHz MOJAVE-5 experiments is approximately a factor of
1.5 greater than derived from the contemporary dual-band 2.3/8.6~GHz
dataset.

\begin{table}
   \caption{Parameters of the fitted linear regression model of baseline length
   repeatability in a form of $a\cdot L + b$ where $L$ is length of baseline in mm.}
   \label{tab:bsl}       
   \begin{tabular}{llrr}
      \hline\noalign{\smallskip}
      dataset & software & $a$ [ppb] & $b$ [mm]\\
      \hline\noalign{\smallskip}
      MOJAVE bl229 & PIMA, pSolve & 0.91 & 2.50 \\
      RV\&CN VLBA only&PIMA, pSolve & 0.64 & 1.51 \\
      RV\&CN all stat&PIMA, pSolve & 0.61 & 1.54 \\
      \hline\noalign{\smallskip}
      MOJAVE bl229 & PIMA, VieVS & 0.98 & 2.04 \\
      RV\&CN all stat & Fourfit, VieVS & 0.60 & 1.17 \\
      \noalign{\smallskip}\hline
   \end{tabular}
\end{table}

\paragraph{Earth orientation parameters}
\label{sec:eop}
The Earth orientation parameters were estimated in a so-called backward solution, i.e. in a solution consistent
with globally estimated terrestrial and celestial reference frames from the processed sessions.
The orientation and the origin of the TRF is set to have no-net-translation (NNT) and no-net-rotation
(NNR) with respect to the ITRF2014 \citep{Altamimi16} for positions of all ten VLBA stations, and the CRF is oriented by imposing
the no-net-rotation condition with respect to ICRF3 \citep{Charlot20} coordinates of defining sources.
We ran several solutions similar to those we introduced in the previous paragraph
and we computed the EOP using both software packages pSolve and VieVS. Table~\ref{tab:eop} shows the wrms of the ERP (polar motion components and dUT1) w.r.t. IERS 14 C04 time series after a trend and bias removal, whereas the wrms of the nutation offsets is given w.r.t. a harmonic expansion heo\_20200606.heo
produced from analysis of available geodetic VLBI data since 1980 through 2020 using the method presented in \citet{r:erm}.
In addition, the median formal error for all five EOP is summarized in the table. We show three solutions computed with pSolve similar to those introduced by the baseline length repeatability, i.e. EOP from MOJAVE-5 dataset, EOP from RV\&CN sessions including all stations, and EOP from RV\&CN sessions using the VLBA telescopes only.
Estimation of EOP using single band observations at high frequencies was made in
the past \citep[e.g.,][]{r:vgaps}. Our recent processing
of 37~VLBA experiments at 24~GHz \citep{Krasna19} showed that
although formal uncertainties were on par with dual-band
regular geodetic experiments (60~$\mu$as for x-pole, 80~$\mu$as
for y-pole and 5~$\mu$s for UT1), the wrms of the difference with respect
to the IERS 14 C04 time series taken as a reference were greater
than formal uncertainties by a factor of three for polar motion and a factor
of ten for UT1. Table~\ref{t:eop} shows that ERP determined from MOJAVE-5
data have the wrms differences with respect to the reference
IERS~C04~14 by a factor of 1.3 to 1.8 larger than from RV\&CN
experiments at the same network.

\begin{table*}
\caption{Wrms and median formal error statistics of the estimated EOP from MOJAVE-5
         bl229 and RV\&CN series. The values for ERP are given w.r.t. IERS 14 C04 time
         series after trend and bias removal. The celestial pole offsets differences dX and
         dY were  computed with respect to the empirical harmonic expansion heo\_20200606.heo.
}
\label{tab:eop}
\begin{tabular}{llrrrrr}
   \hline\noalign{\smallskip}
      && x-pole [$\mu$as]& y-pole [$\mu$as] & dUT1 [$\mu$s] & dX [$\mu$as]& dY [$\mu$as]\\
      \noalign{\smallskip}\hline\noalign{\smallskip}
        MOJAVE bl229 & wrms   & 228 & 286 & 23 & 169 & 128 \\ \vspace{2ex}
      & median formal error& 109 & 153 & 9 & 59 & 56 \\
      RV\&CN VLBA only & wrms    & 126 & 218 & 15 & 89 & 129 \\ \vspace{2ex}
      & median formal error&  80 & 120 & 6 & 92 & 69 \\
      RV\&CN all stat & wrms   & 117 & 130 & 14 & 72 & 87 \\
      & median formal error &  57 &  89 & 4 & 86 & 60 \\
      \noalign{\smallskip}\hline\noalign{\smallskip}
      \label{t:eop}
\end{tabular}
\end{table*}

\begin{table*}
     \caption{Mean number of scans at VLBA telescopes in one session computed over the period of interest (September 2016 - July 2020).}
     \label{tab:nrscans}       
     \begin{tabular}{lllllllllll}
          \hline\noalign{\smallskip}
          & Br & Fd & Hn & Kp & La & Mk & Nl & Ov & Pt & Sc\\
          \noalign{\smallskip}\hline\noalign{\smallskip}
          MOJAVE bl229 series & 245 & 245 & 241 & 248 & 251 & 204 & 251 & 252 & 235 & 219\\
          geodetic RV\&CN experiments & 451 & 485 & 445 & 493 & 483 & 357 & 467 & 487 & 451 & 423\\
     \noalign{\smallskip}\hline
     \end{tabular}
\end{table*}

\section{Differences between MOJAVE-5 bl229 and RV\&CN}

We recognize there are three major differences between the
datasets which may have an impact on geodetic results. First,
different scheduling approaches were used due to different goals
of the experiments. Second, modeling of the ionospheric path delay
was different since MOJAVE-5 was observed at a single-band.
Third, different radio sources were selected for observations.
We isolate these factors and determine which factor
has the greatest impact on the accuracy of geodetic solutions.

\subsection{Scheduling}

\begin{figure*}
  \includegraphics[width=0.33\textwidth]{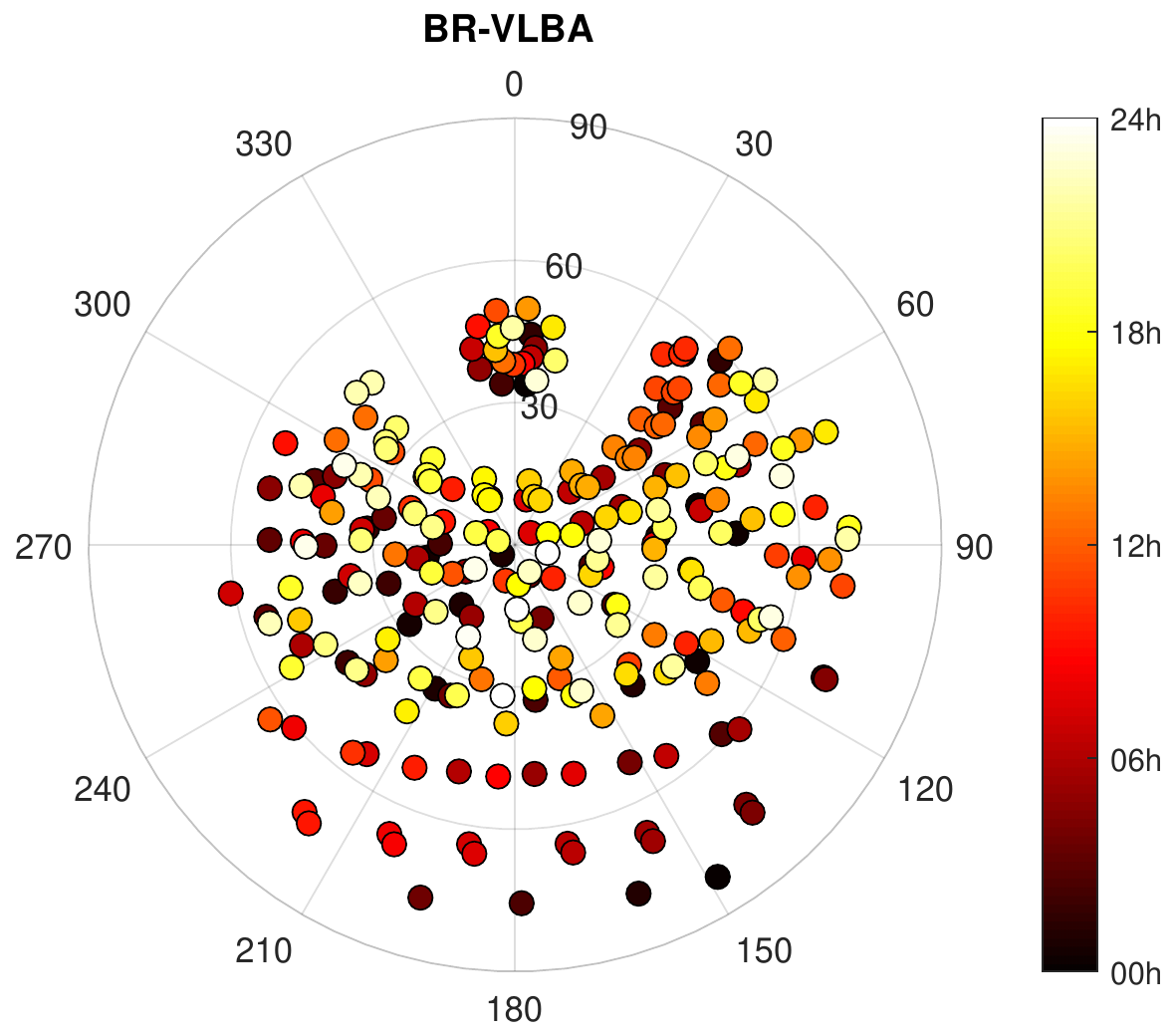}
  \includegraphics[width=0.33\textwidth]{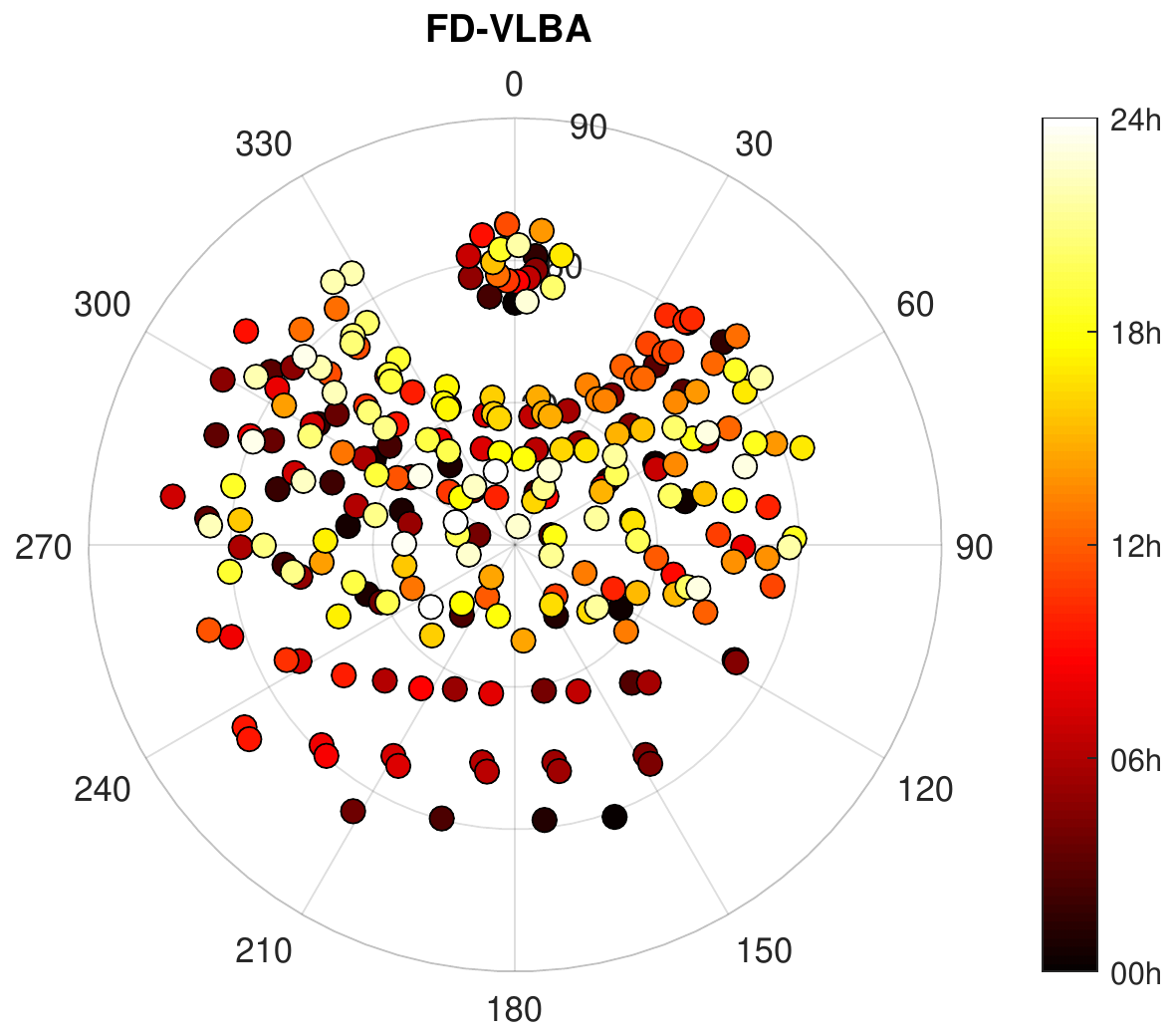}
  \includegraphics[width=0.33\textwidth]{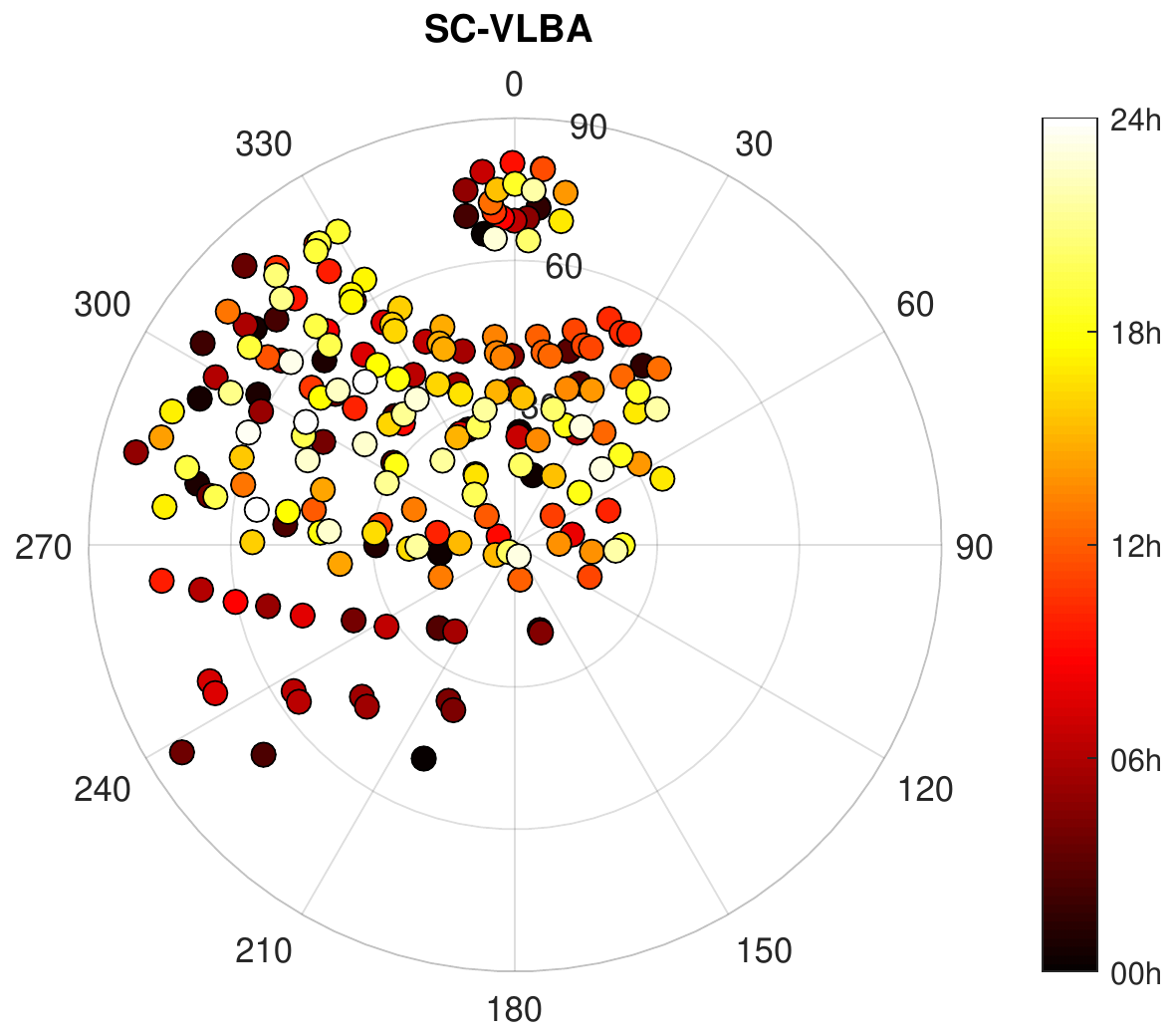}\\
  \includegraphics[width=0.33\textwidth]{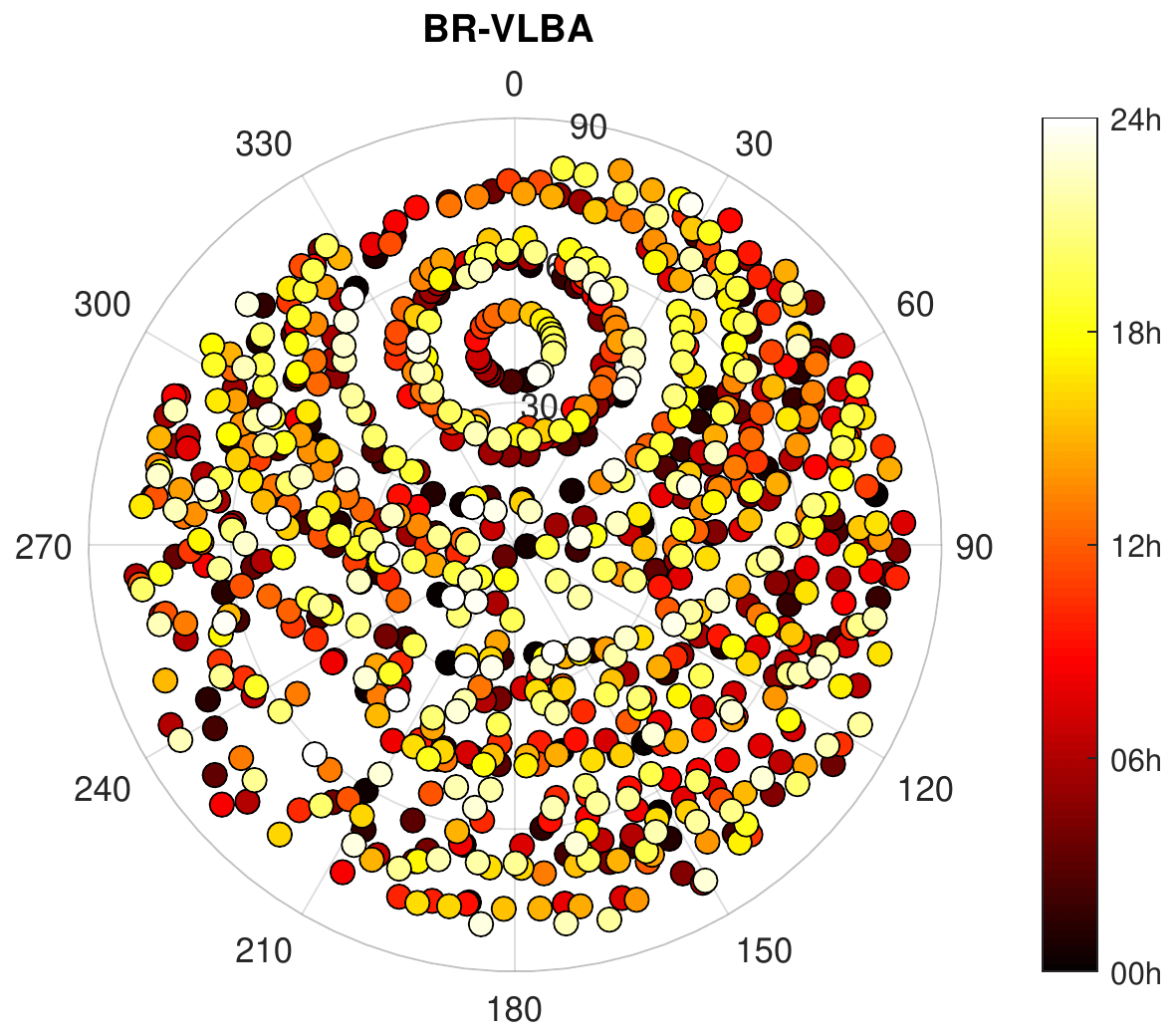}
  \includegraphics[width=0.33\textwidth]{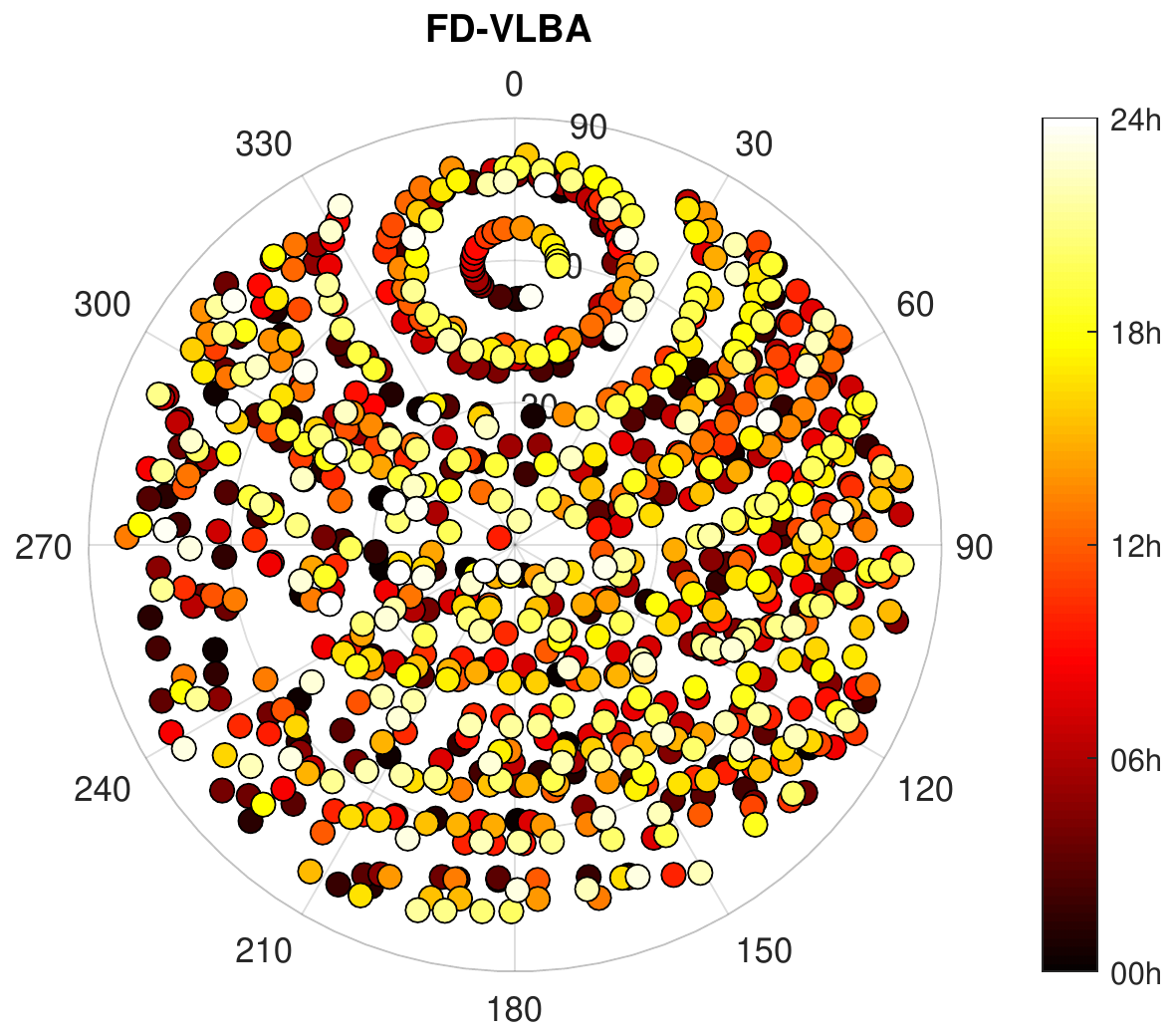}
  \includegraphics[width=0.33\textwidth]{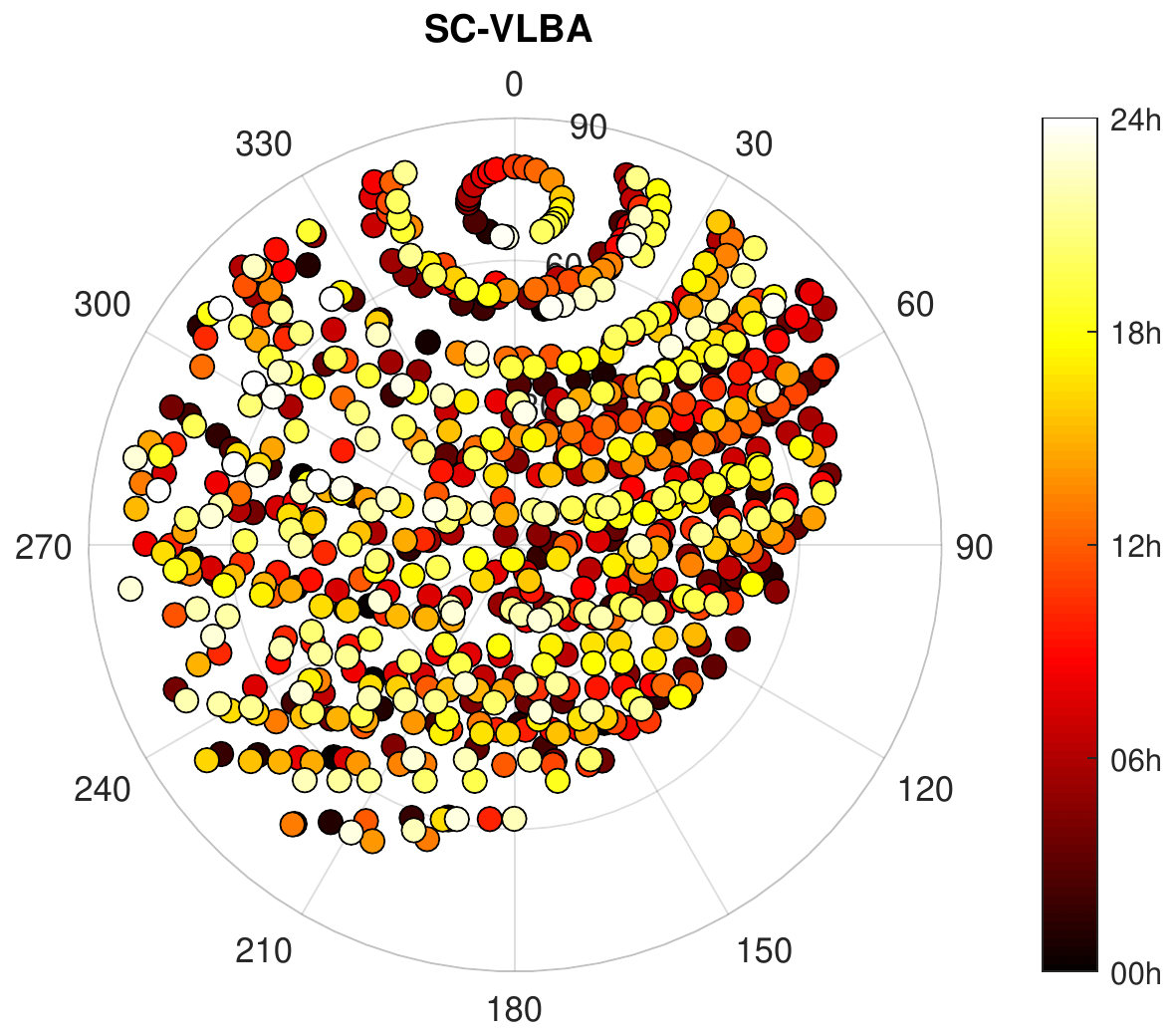}
\caption{Sky coverage of three VLBA stations: BR-VLBA, FD-VLBA and SC-VLBA during the bl229bc MOJAVE-5 experiment (upper plots) and the cn1924 experiment (lower plots). }
\label{fig:skycov}       
\end{figure*}

\begin{figure}
  \includegraphics[width=\hsize]{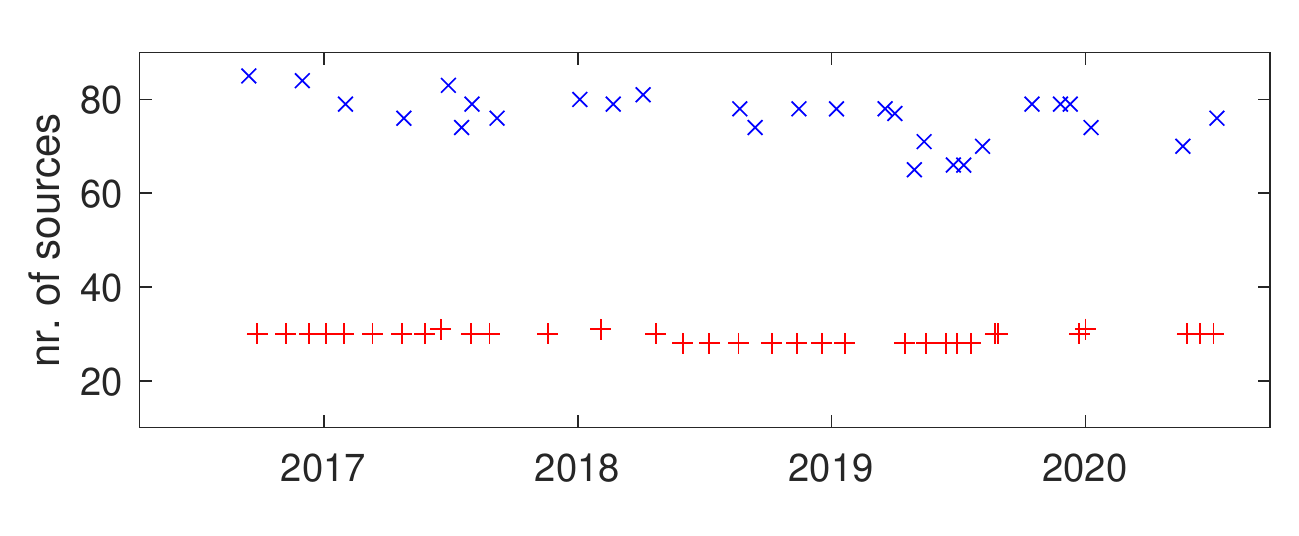}\\
  \includegraphics[width=\hsize]{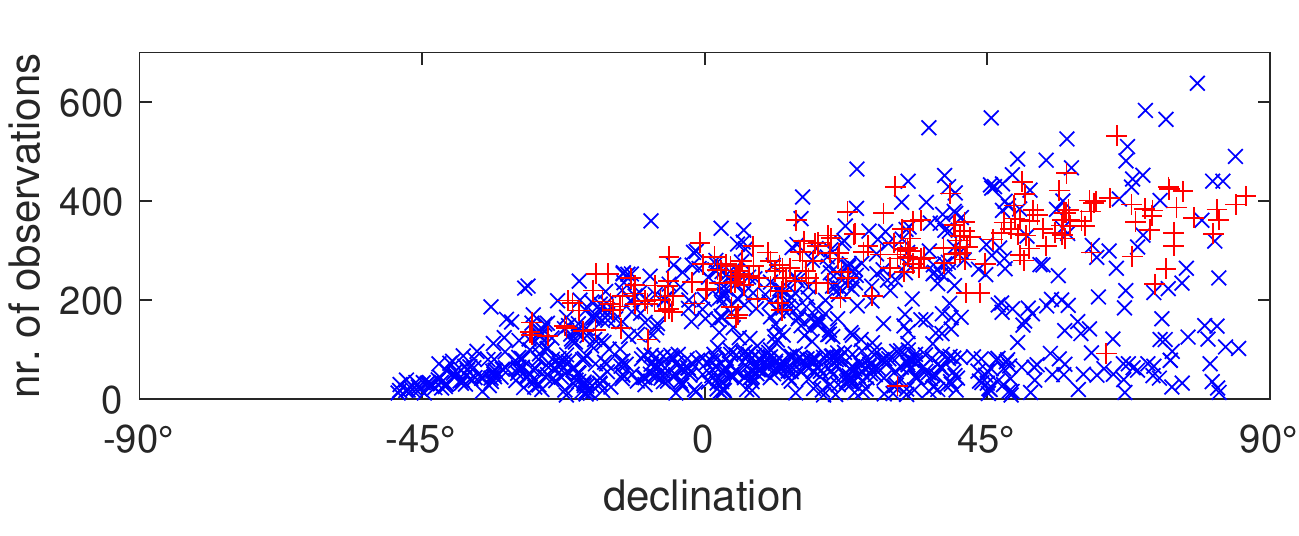}
\caption{The upper plot shows number of observed sources in each session. The lower plot depicts the median number of observations for each source. The red crosses stand for the bl229 experiments, blue x-signs for the RV\&CN experiments.}
\label{fig:nrsou}       
\end{figure}

Scheduling of a VLBI experiment is a complex task. The scheduler has to evaluate several criteria which lead to the best schedule according to the focus of the current experiment.\\
The major criteria for design of the geodetic schedule are the sky coverage over the individual stations, number of observations, and scan duration. The general scheduling concept of the established software sked for the geodetic sessions can be found in \citet{Gipson10}.
Evenly distributed observations over all elevation angles at a given station ensure a good
decorrelation of station dependent parameters such as station height, zenith wet delay, clock parameters, or baseline clock offsets (e.g., \citet{Nothnagel02}), and therefore, such schedule can be regarded as station-centric. A large number of observations in general improves the accuracy of the estimated geodetic parameters due to higher redundancy. The challenge for the scheduling software is to find the best compromise between a) the long antenna slew time needed for the best sky coverange in a short interval (1-3 hours) allowing a high time resolution of the estimated parameters and b) the short antenna slew time allowing for a high number of observations with sufficient scan duration and high signal to noise ratio. The newly developed scheduling software VieSched++ \citep{Schartner19,Schartner20} written in modern C++ which is a successor of a Matlab based VieVS scheduler \citep{Sun13,Sun14} provides sophisticated optimization criteria and allows to create and select schedules based on simulations of the proposed observations. Recently, a new geodetic scheduling approach inspired by evolutionary processes was presented by \citet{Schartner21} which enables the generation of fully automated and individually optimized schedules providing up to 10\% more observations compared to previous scheduling approaches.\\
Astronomic VLBI schedules are mainly scheduled using Sched \citep{Walker20}. The primary goal of
the MOJAVE-5 bl229 experiments is to provide best images of jets in active galactic
nuclei. Therefore, their schedule is optimized to track a small set of sources (30)
in a 24-hour session in ten scans per source for a total on-source time of $\sim35$ minutes. A given source is scheduled to have at least 6 antennas above $10^\circ$ elevation\footnote{\href{http://www.physics.purdue.edu/astro/MOJAVE/data.html}{http://www.physics.purdue.edu/astro/MOJAVE/data.html}}. Such schedule is considered as source-centric.\\
In Fig.~\ref{fig:skycov} we show the sky coverage during a 24-hour
observing session at three selected telescopes (BR-VLBA, FD-VLBA, SC-VLBA)
where colors depict the time passed since the session start. As an example,
we show the sky coverage during the MOJAVE-5 session bl229bc observed on December 22, 2019 in the upper plots and the cn1924 session observed with the same network on December 09, 2019 in the lower plots. Table~\ref{tab:nrscans} summarizes the mean number of scans in a 24-hour experiment at each of the ten VLBA telescopes computed over the investigated
time period (September 2016 --- July 2020). We see that twice as many scans
at each telescope were observed in geodetic experiments with short integration
time than in MOJAVE-5 observing sessions that used longer integrations.
Fig.~\ref{fig:nrsou} presents the total number of observed sources in each
session (upper plot) and the median number of observations during a 24-hour session for each source computed over the respective four years period.
The median number of observed radio sources is 30 in MOJAVE-5 sessions, and 78 in RV\&CN sessions.
Comparison of the number of observations for each source during a whole session shows that 95\% of the AGNs
observed in MOJAVE-5 sessions have more than 150~observations, whereas only 35\%
of the sources were observed in RV\&CN that often. It approves that geodetic schedules
are designed to provide a good sky coverage for each station and the sources are
picked up to improve the azimuth/elevation coverage regardless of how often
they are observed in a given experiment.

\subsection{Ionosphere}
\label{sec:iono}

  The ionosphere is a refractive media. Propagating in the ionosphere, phase
delay decreases and group delay $\tau_{gr}$ increases with respect to
the ionosphere free $\tau_{if}$ group delay in the absence of the ionosphere as
\begin{equation}
   \tau_{gr} = \tau_{if} + \kappa{\rm \Delta {\rm TEC}}/f^2_{\rm eff} ,
   \label{e:e1}
\end{equation}
   where $f_{\rm eff}$ is the effective frequency that is within several
percent of the recorded central sky frequency, $\Delta$TEC is the differential
Total Electron Content measured in TEC units (TECU, 1 TECU = $10^{16}$ electron/$m^2$):
\begin{equation}
      \Delta {\rm TEC} = \int N_v \, d s_1 - \int N_v \, d s_2
\end{equation}
  with $s_1$ and $s_2$ as paths of wave propagation from a source to the
first and second station of the radio interferometer, and
\begin{equation}
 \kappa = 10^{-16} \cdot \frac{e^2}{ 2 \, \textup{c} \, m_e \,  \epsilon_o} = 5.308018 \cdot 10^{10} \; \textup{s}^{-1}
\end{equation}
where $e$ --- charge of an electron, $m_e$ --- mass of an electron,
$\epsilon_o$ --- permittivity of free space, c --- velocity
of light in vacuum.

  To mitigate the impact of the ionosphere on group delay, geodetic observations
are usually conducted at two frequencies simultaneously. Combining group delays
$\tau_u$ and $\tau_l$ at the upper and lower frequencies $f_u$ and $f_l$
respectively, we can derive the differential TEC and the ionosphere free path
delay as

\begin{equation}
   \begin{array}{lcl}
      \Delta TEC & = & \Frac{f^2_u   f^2_l}{f^2_u - f^2_l} \; (\tau_l - \tau_u), \\
      \tau_{if}  & = & \Frac{f^2_u}{f^2_u - f^2_l} \, \tau_u -
                       \Frac{f^2_l}{f^2_u - f^2_l} \, \tau_l  , \\
      \tau_{iu}  & = & \Frac{f^2_l}{f^2_u - f^2_l} \; (\tau_l - \tau_u).
   \end{array}
\end{equation}

  Derivations of these equations can be found for example in \citet{Petrov11}. This approach allows to effectively cancel the ionospheric contribution,
leaving residual contribution at a level not exceeding several picoseconds \citep{r:hob05}.

  MOJAVE-5 program used only one frequency. An alternative approach for
modeling the ionospheric contribution is to use TEC maps from
GNSS observation processing. Applying time and spacial interpolation, we can
compute TEC in the vertical direction for each station and each observation.
Then we can relate the TEC in the direction of observation at the elevation
angle $E$ to the TEC in the vertical direction via a mapping function $M_i(E)$.
Considering the ionosphere as a thin shell at height $H$, we can easily
derive the ionospheric mapping function as
\begin{equation}
   \begin{array}{lcl}
    M_i(E)  & = & \Frac{1}{\cos{\beta(E)}},\\
     \beta(E) & = & \arcsin \Frac{ \cos E }{1 + \frac{H}{R_\oplus}} \vspace{0.5ex}, \\
   \end{array}
\end{equation}
    where $R_\oplus$ is the Earth's radius.

  We used Center for Orbit Determination in Europe (CODE) TEC time series
\citep{r:schaer99}\footnote{Available at
\href{ftp://ftp.aiub.unibe.ch/CODE}{ftp://ftp.aiub.unibe.ch/CODE}} with
a resolution of $5^\circ \times 2.5^\circ \times 2^h$. This resolution
is relatively coarse and accounts only for a part of the signal. Therefore,
our results of processing MOJAVE-5 observations are affected by systematic errors
caused by the residual ionosphere.

  In order to quantify the residual ionospheric signal, we processed dual-band
RV\&CN data set. For the purpose of this study, we consider that the ionospheric
free linear combination of X and S band group delays has no ionospheric
contribution. We can form the differences between the ionospheric contribution
computed from TEC maps and from X and S band group delays and investigate the
properties of this stochastic process.

\begin{figure}
  \includegraphics[width=0.484\textwidth]{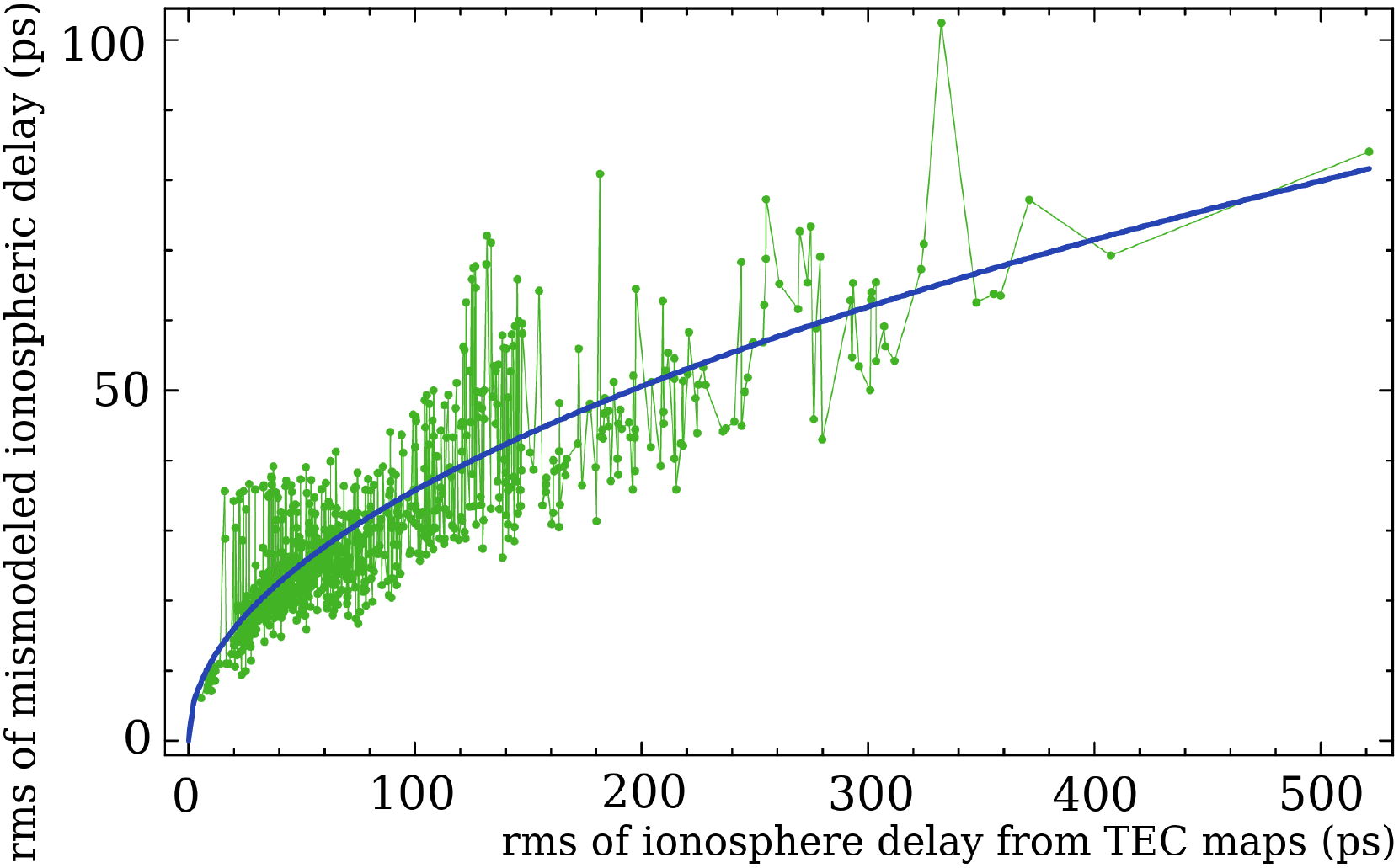}
  \caption{The rms of the errors in the ionospheric
           path delay as a function of the rms of
           the variations of the ionospheric group delays derived
           from TEC maps (green dots).
           The solid blue line shows a regression in
           a form of the power law 1/2.
          }
  \label{f:iono_rms_mod}
\end{figure}

  Solving for zenith path delay in the neutral atmosphere will pick up
a portion of the slowly varying bias, but the ionospheric fluctuations
at scales less than several hours will propagate to residuals. Though,
we can characterize stochastic properties of the residual signal similar
to the approaches developed in \citet{r:lcs1,r:lcs2,r:wfcs}.
The ionospheric path delay fluctuation is a non-stationary process.
We can expect that fluctuations at scales $x$ will be related to fluctuations
at scales $y$ via a power law from the general results of the turbulence theory.
Therefore, we did the following:

  First, we computed the mean differences of $d_{gv} = \tau_{ig} - \tau_{iv}$
between the ionospheric path delay at X band computed from TEC maps ($\tau_{ig}$)
and from VLBI dual band observables ($\tau_{iv}$) for every baseline and every
experiment in the RV\&CN dual-band dataset, and then we subtracted the mean value
from $d_{gv}$. The mean value is the sum of the bias between TEC maps and VLBI
ionospheric path delay and a constant instrumental delay in VLBI hardware. Since
the constant instrumental delay that may be even greater than the ionospheric signal
is not calibrated, the mean value of $d_{gv}$ is meaningless. Then we computed
the rms over $d_{gv}$. We discarded the data with clock jumps that may happen
at only one band. We got time series of rms($d_{gv}$) and we examined empirical
relationships of rms($d_{gv}$) with other statistics. We found that
rms${}^2$($d_{gv}$) has a linear dependence on rms($\tau_{ig}$). The power
law dependence between $d_{gv}$ and $\tau_{ig}$ was expected, but the power
law coefficient, 2, is purely empirical. Fig.~\ref{f:iono_rms_mod}
demonstrates the time series of $d_{gv}$ and their fit.

    We can compute the rms of the ionospheric errors at a given baseline
of a given experiment via
\begin{equation}
   {\rm rms}(d_{gv}) = \sqrt{ \rho \; {\rm rms}(\tau_{ig})},
   \label{e:e2}
\end{equation}
  where $\rho$ is the empirical coefficient determined from fitting (see Fig.~\ref{f:iono_rms_mod})
equal to 12.8~ps and the rms is expressed in ps. This empirical relationship allows us
to predict the second moment of the residual noise after we perform
data reduction for the ionospheric contribution using TEC maps.
One can expect that if the TEC variance is greater, the residual errors
are also greater. Expression~(\ref{e:e2}) quantifies this dependence.

\begin{figure}
  \includegraphics[width=0.5\textwidth]{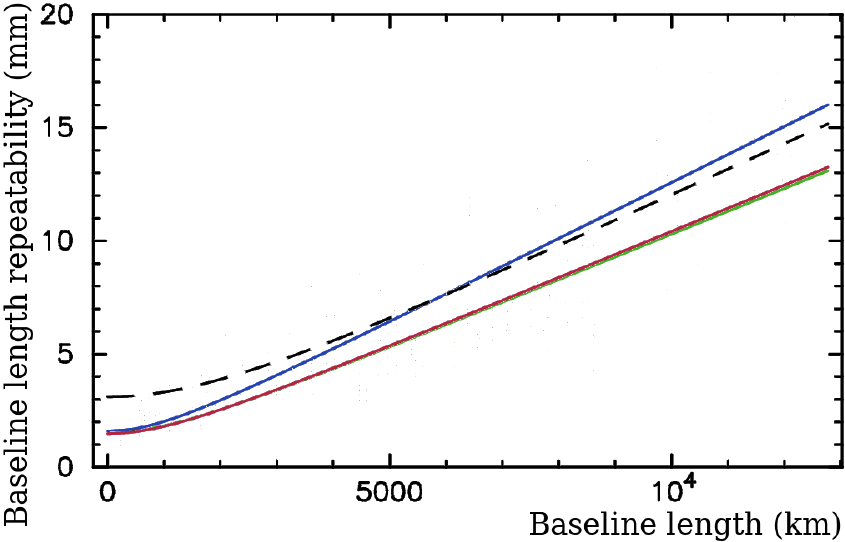}
  \caption{The dependencies of the baseline length repeatability fits on
           the baseline length. The upper blue curve shows the baseline
           repeatability for the X band only in the ``bx'' solution that uses
           GNSS TEC maps. Two lower very close curves, red and green,
           show the baseline length repeatability for the ``bu'' solution
           that demonstrates the effect of mismodeled ionosphere on Ku band
           observable, and the reference dual band solution. The dashed black
           line shows the baseline length repeatability from the MOJAVE-5
           solution.
          }
  \label{f:iono_basrep}
\end{figure}

  We have computed baseline-dependent additive noise due to mismodeled
ionosphere for every baseline and every experiment of MOJAVE-5 program
using $\tau_{ig}$. We added that noise to the a priori group delay errors
in quadrature and computed new weights. We ran several baseline solutions,
computed baseline repeatabilities, and compared them with the reference
dual-band solution using RV\&CN data. In solution ``bx'' we used the
ionosphere-free combinations of group delays, added the contribution of
the ionosphere $\tau_{iu}$ to them, and processed these data the same way
as MOJAVE-5 data, i.e. performing data reduction for the ionosphere using
CODE TEC maps and inflating a priori group delay uncertainties for the
additional noise due to mismodeling the ionosphere. In the second solution
``bu'' we simulated how the deficiency of CODE TEC model would have impacted
our RV\&CN solution, as if these experiments ran at 15.3~GHz instead of
2.3/8.6~GHz. To achieve this, we re-scaled $\tau_{iu}$ by the square of
the frequency ratio $(8.64/15.28)^2 \approx 0.32$. Fig.~\ref{f:iono_basrep}
shows fit in a form $\sqrt{(a\,L)^2 + b^2}$ for all these solutions. The baseline
length repeatability from MOJAVE-5 solution is shown by the dashed line.

  We found that the impact of the mismodeling ionosphere on the baseline
length repeatability of VLBA data collected in 2016--2020 at 15.3~GHz during
Solar minimum is negligible. Therefore, an increase in the baseline length
repeatability from a geodetic solution using the MOJAVE-5 dataset with
respect to the reference dual-band RV\&CN solution cannot be explained
by the unaccounted contribution of the ionosphere. We should caution
that this result should not be extrapolated to other estimated parameter,
such as source position, and should not be extrapolated to epochs
of the Solar maximum.

\subsection{Simulations}

To touch the effect of source structure, we run simulations of the
observations. Using the simulation VieVS tool \citep{Pany11}, we
replaced group delay observables with synthetic artificial group
delays provided by the random noise generator as
\begin{equation}
   \tau_{gr} = \tau_{mod} \:\:\: + \:\:\: (\tau_{clk} + \tau_{zwd} + \tau_{fl}).
   \label{e:sto}
\end{equation}

We add three stochastic error sources to the theoretically computed time delay ($\tau_{mod}$): delay caused by the turbulence in the troposphere ($\tau_{zwd}$), station clock ($\tau_{clk}$), and the Gaussian noise ($\tau_{fl}$) with $\sigma$=20~ps that accounts for the thermal noise and instrumental errors.
We used the model of \citet{Nilsson07} for simulation of zenith wet delay implemented in VieVS. In the framework of that approach we consider that the atmospheric turbulence for every station is described with a structure function with the refractive index structure constant $C_n = 1.8 \cdot 10^{-7} \; \textup{m}^{-1/3}$, the effective height $H = 2~\textup{km}$, and the constant wind velocity $v =$ 8~m/s towards East. Then we computed the covariance matrix between group delays for each pair of observations of a given station and used them for computation of the full weight matrices under an assumption that the atmospheric turbulence is a stationary process. The simulation of station clocks was performed with an Allan standard deviation of $1\cdot10^{-14}$ at 50~min. \strike{, and we applied the white noise as the Gaussian process with $\sigma=20$~ps.} We did not include modeling source structure into simulation.

\begin{figure}
  \includegraphics[width=0.5\textwidth]{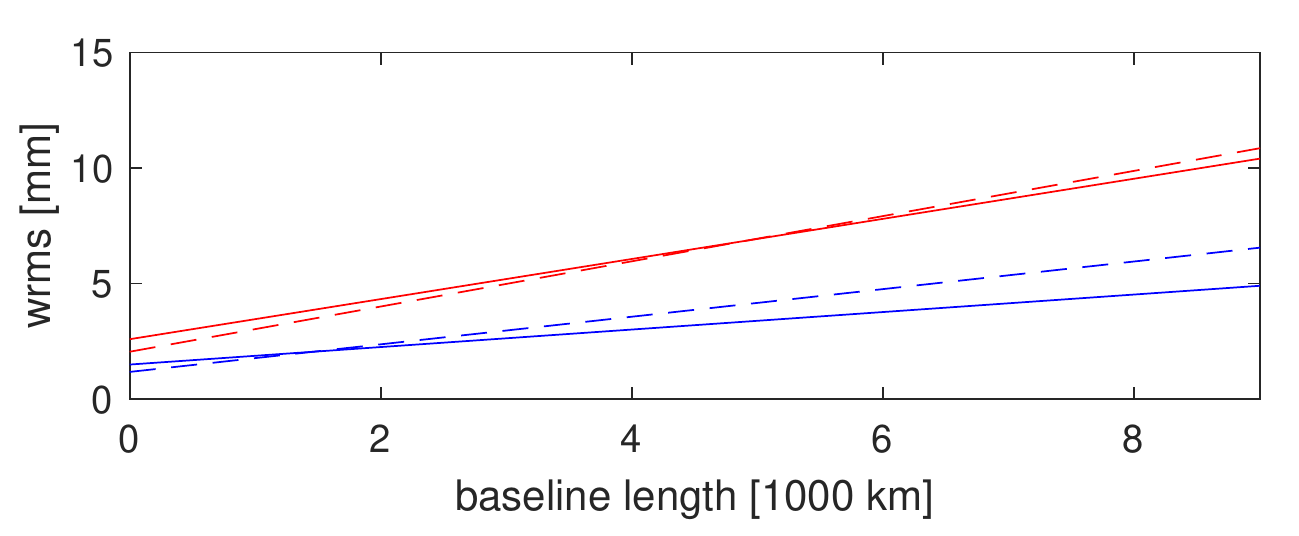}
  \caption{The wrms of baseline length from simulated
           (dashed line) and real observations (solid line).
           The upper red lines show the baseline length repeatability
           from analysis and simulation of MOJAVE-5 data. The low blue
           lines shows results of analysis and simulation of RV\&CN data.
          }
  \label{fig:bslsim}
\end{figure}

  We have computed baseline length repeatabilities from simulated RV\&CN and
MOJAVE-5 datasets. The regression lines in a form $a\cdot L + b$ for simulated and real data are shown in Fig.~\ref{fig:bslsim}. We see from these plots that simulation results show even deeper disparity in repeatabilities between MOJAVE-5 and RV\&CN data as we saw in real observations.
We should note that $\tau_{clk}$ and $\tau_{fl}$ in equation (\ref{e:sto}) are exactly the same for both datasets, and $\tau_{zwd}$ that is dependent on elevations and time is similar in both datasets because it is derived from the same model.
Therefore, the stochastic model we used for MOJAVE-5 and RV\&CN simulations is essentially the same. This finding pinpoints the origin of discrepancies in results of processing real data: differences in schedules.

\begin{figure*}
  \includegraphics[width=0.499\textwidth]{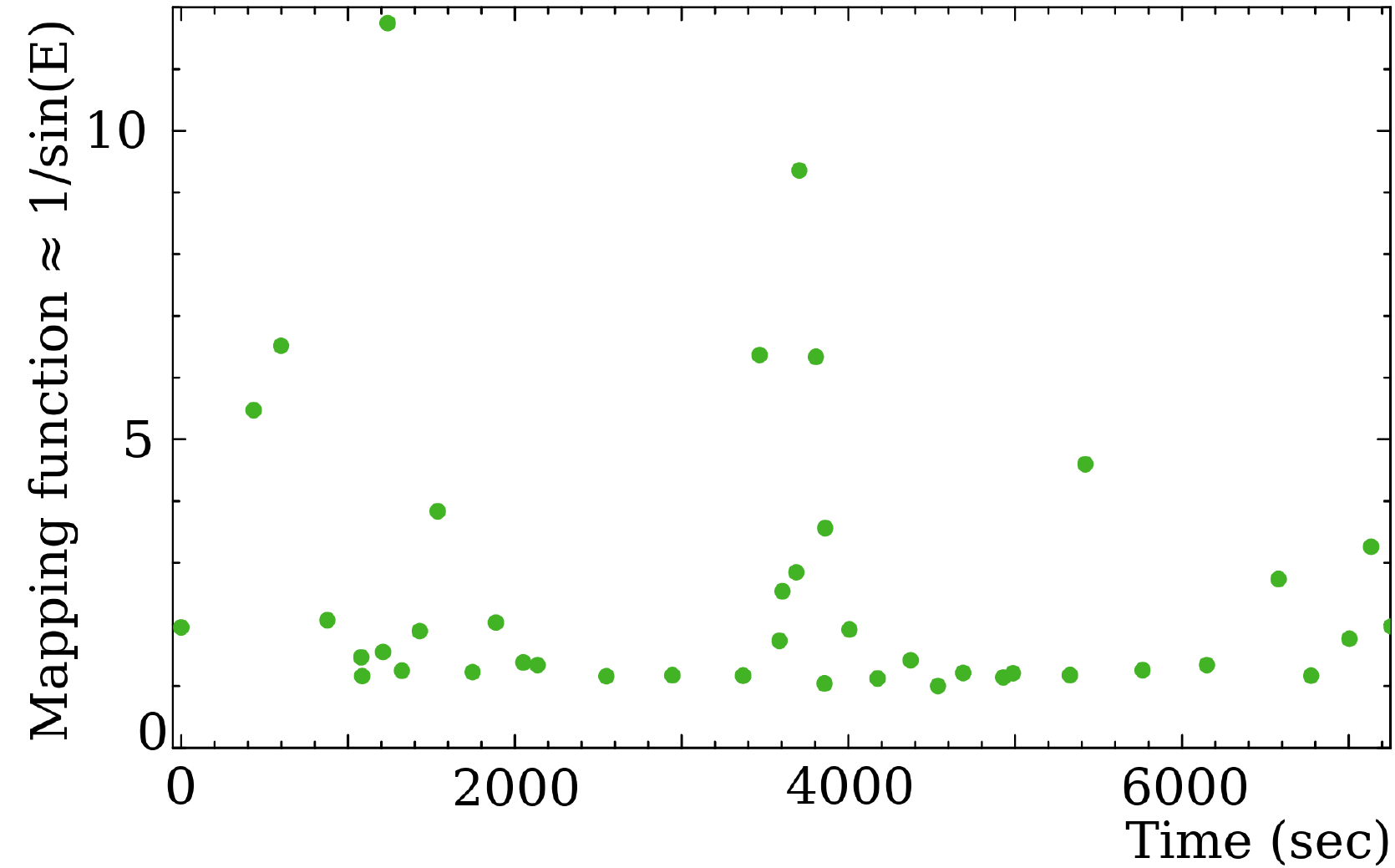}
  \hspace{0.001\textwidth}
  \includegraphics[width=0.499\textwidth]{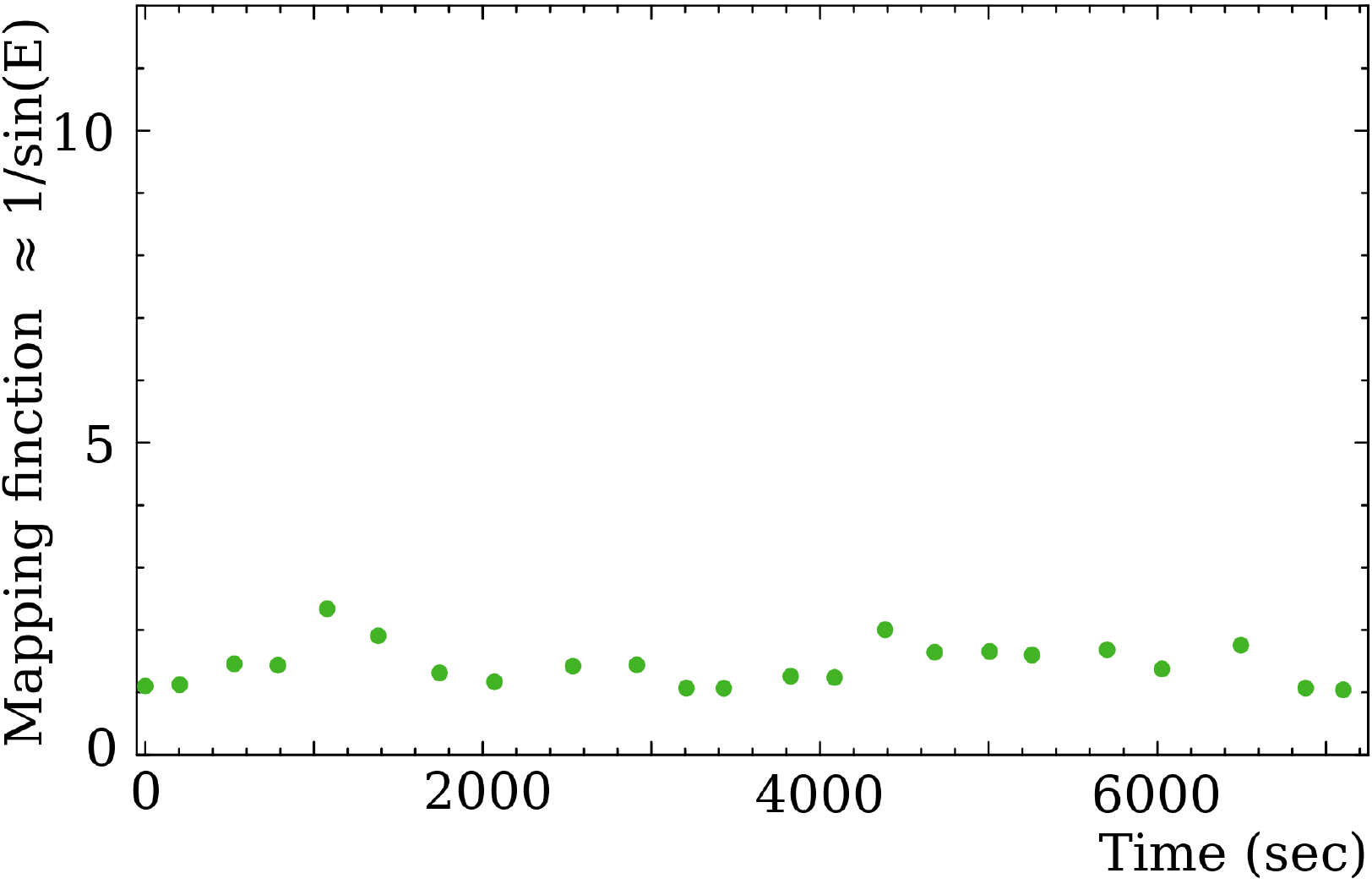}
  \caption{The distribution of observations over the mapping function for station
           LA-VLBA within first two hours of an experiment. Since deviation
           of the mapping function from 1/sin(elevation) is small at elevations
           above 10 deg, mapping function 1/sin(elevation) is used here for
           illustrative purpose.
           {\it Left:  } geodetic experiment rv119.
           {\it Right: } astronomical experiment bl229aa.
          }
  \label{f:eltim}
\end{figure*}

 Since atmospheric
path delay and clock function are modeled in a form of an expansion
over the B-spline basis with a time span 20--60~minutes, in order to
decorrelate these two groups of nuisance parameters and the station vertical component,
observations at significantly different elevations are required.
Fig.~\ref{f:eltim} shows that the spread of observations over
mapping function (approximately reciprocal to sine of elevation angle which equals to the partial derivative of the time delay w.r.t. zenith wet delay)
for geodetic experiment rv119 is noticeably wider and observations at
elevations below $30^\circ$, which corresponds to mapping function $>2$,
are much more often than in astronomical experiment bl229aa.

\begin{table}
   \caption{The wrms of the simulated EOP from MOJAVE-5 bl229 and RV\&CN
            experiments observed at the VLBA network. The wrms statistics
            of the differences between EOP estimates the IERS 14 C04 time
            series after trend and bias removal are shown below.
           }
   \par\hspace{-2ex}\par
   \begin{tabular}{l@{}rrrrr}
           \hline
           &  x-pole  & y-pole  & dUT1   & dX      & dY          \\
           &  [$\mu$as] & [$\mu$as] & [$\mu$s] & [$\mu$as] & [$\mu$as]     \\
           \hline
           MOJAVE-5 bl229   & 158 & 225 & 16 & 82 & 98 \\
           RV\&CN VLBA only & 118 & 184 & 10 & 82 & 97 \\
           \hline
    \end{tabular}
    \label{tab:sim_eop}
\end{table}

\begin{figure}
   \includegraphics[width=0.5\textwidth]{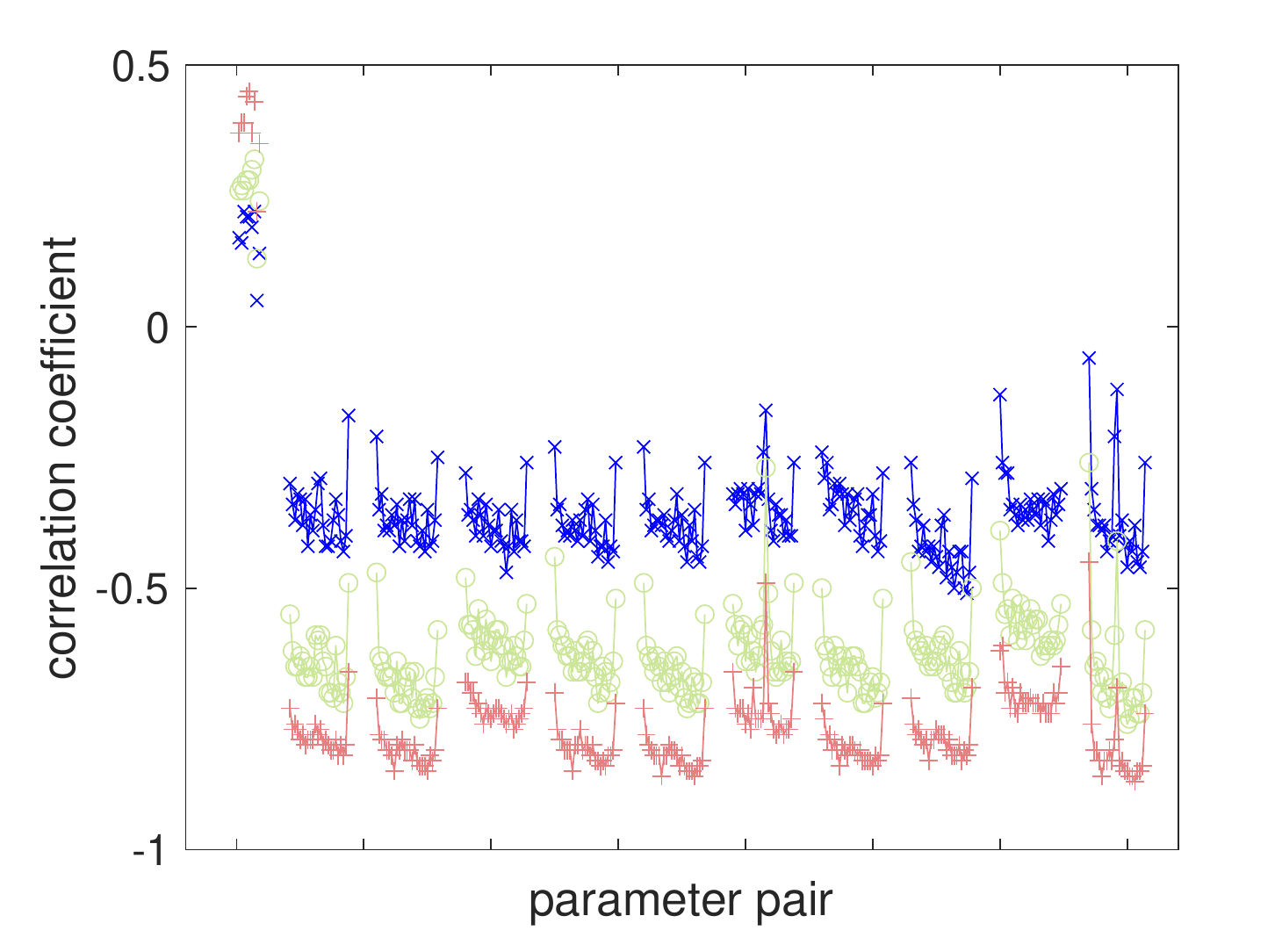}
   \caption{Correlation coefficients in simulated geodetic
           experiment cn1924 for cutoff elevation 3$^\circ$ (blue x-signs), 20$^\circ$ (green circles)
           and~30$^\circ$ (pink crosses). The order of parameter pairs at the
           x axis: (1st--9th): station's vertical component and clock offset for all
           stations except of the clock reference; (10th--last): station's vertical
           component and a residual atmospheric zenith path delay (estimated as pwlo every hour) for all ten stations
           (the solid line connects coefficients which belong to the same station).
           }
   \label{fig:elev-cutoff}
\end{figure}

In order to look  at the problem deeper, we investigated correlations
between estimates of the vertical site position and atmospheric
path delays. The median correlation coefficient is -0.22 for RV\&CN sessions and
-0.35 for MOJAVE-5 sessions computed from the real data. We ran a series of solutions using RV\&CN data
and flagged out observations below a certain elevation angle.
Table~\ref{tab:elev_cutoff} summarizes the findings. An increase of
the elevation cutoff results in an increase of the baseline length
repeatability. MOJAVE-5 has few scans below elevations $30^\circ$ and none
below $20^\circ$. The achieved baseline length repeatability from
MOJAVE-5 experiments is similar to the repeatability from RV\&CN experiments
when observations below 20--$25^\circ$ are not included in a solution.
Fig.~\ref{fig:elev-cutoff} shows individual correlation coefficients in
simulated geodetic experiment cn1924 for cutoff elevation 3$^\circ$, 20$^\circ$
and 30$^\circ$. The median correlation coefficient between vertical displacement and clock offset at the respective station is 0.19, 0.27 and 0.39 for the increasing cutoff elevation angle.  The median correlation coefficient between vertical component and a residual atmospheric zenith path delay is -0.37, -0.64 and -0.79 when the elevation cutoff is increasing. This proves that strategy including radio sources under low elevations in the schedule over short periods of time allows to decorrelate station dependent parameters in the data analysis and to provide better baseline length repeatability.\\
 We also investigated the residual EOP estimates from the simulated RV\&CN and MOJAVE-5 datasets with respect to the IERS C04 14 time series
taken as a reference. The results of this simulation are presented in
Table~\ref{tab:sim_eop}. The simulation results confirm about the same disparity of 20-60\% of the wrms from MOJAVE-5 and RV\&CN dataset as in the results derived from real observations (compare with Table~\ref{tab:eop}).

\begin{table}
   \caption{The coefficients of the baseline length repeatability
            regression in a form of $a \cdot L + b$ as a function of
            elevation angle from processing RV\&CN VLBA observations.
           }
   \par\hspace{-2ex}\par
   \begin{tabular}{ccc}
           \hline
             Elevation cutoff &  $a$ [ppb]  & $b$ [mm]    \\
           \hline
               $  0^\circ $   & 0.55 & 1.46 \\
               $  5^\circ $   & 0.55 & 1.46 \\
               $ 10^\circ $   & 0.70 & 1.35 \\
               $ 15^\circ $   & 0.90 & 1.14 \\
               $ 20^\circ $   & 1.09 & 1.14 \\
               $ 25^\circ $   & 1.22 & 1.14 \\
               $ 30^\circ $   & 1.86 & 1.49 \\
           \hline
    \end{tabular}
    \label{tab:elev_cutoff}
\end{table}

\subsection{Insight on the possible contribution of the source structure}
\begin{figure*}
  \includegraphics[trim=0cm 0cm 0cm 0cm, clip=true, width=0.33\textwidth]{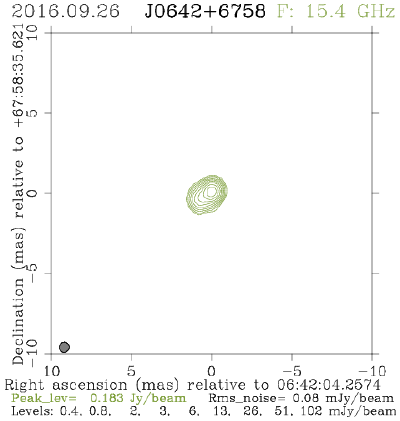}
  \includegraphics[trim=0cm 0cm 0cm 0cm, clip=true, width=0.33\textwidth]{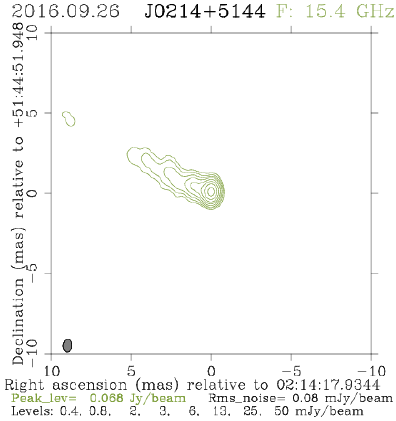}
  \includegraphics[trim=0cm 0cm 0cm 0cm, clip=true, width=0.33\textwidth]{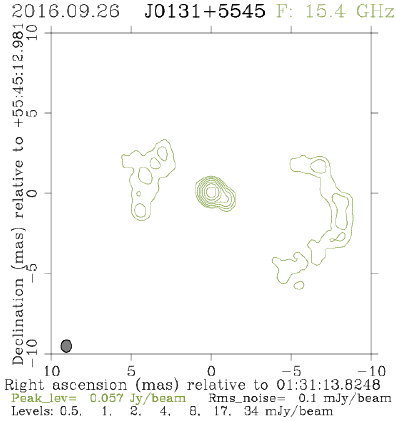}\\
  \includegraphics[trim=0cm 0cm 0cm 0cm, clip=true, width=0.33\textwidth]{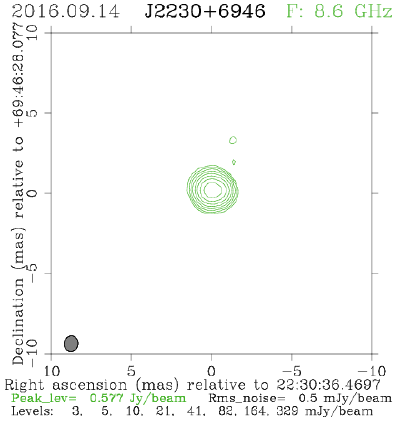}
  \includegraphics[trim=0cm 0cm 0cm 0cm, clip=true, width=0.33\textwidth]{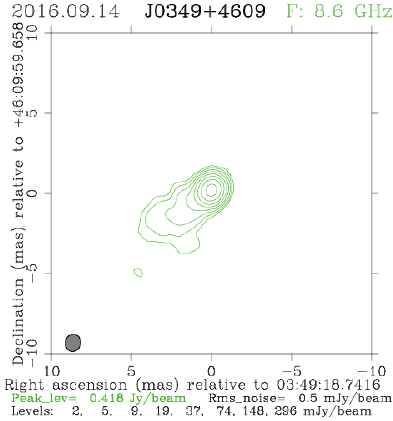}
  \includegraphics[trim=0cm 0cm 0cm 0cm, clip=true, width=0.33\textwidth]{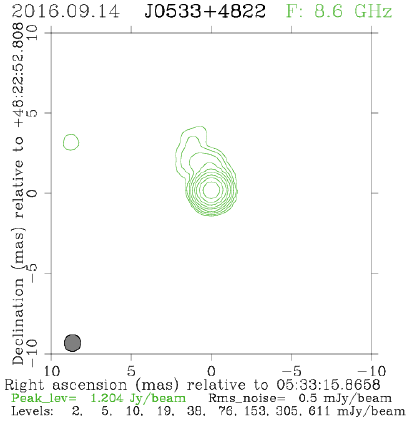}
\caption{Images of three most observed sources (0636+680 with SI 2, 0210+515
with SI 3 and 0128+554 with SI 4) in the bl229aa MOJAVE-5 experiment
(upper plots) and in the rv119 experiment at X band
(lower plots: 2229+695 with SI 2, 0345+460 with SI 2, 0529+483 with SI 2).
We have produced images from rv119 ourselves. The images in FITS format
are available in the Astrogeo VLBI FITS image database
\href{http://asteogeo.org/vlbi\_images}{http://asteogeo.org/vlbi\_images}.
Information about the structure index for the X band sources was taken
from the Bordeaux VLBI Image Database available at
\href{http://bvid.astrophy.u-bordeaux.fr}{http://bvid.astrophy.u-bordeaux.fr}.
}
\label{fig:souimage}       
\end{figure*}

\begin{figure*}
    \includegraphics[width=0.48\textwidth]{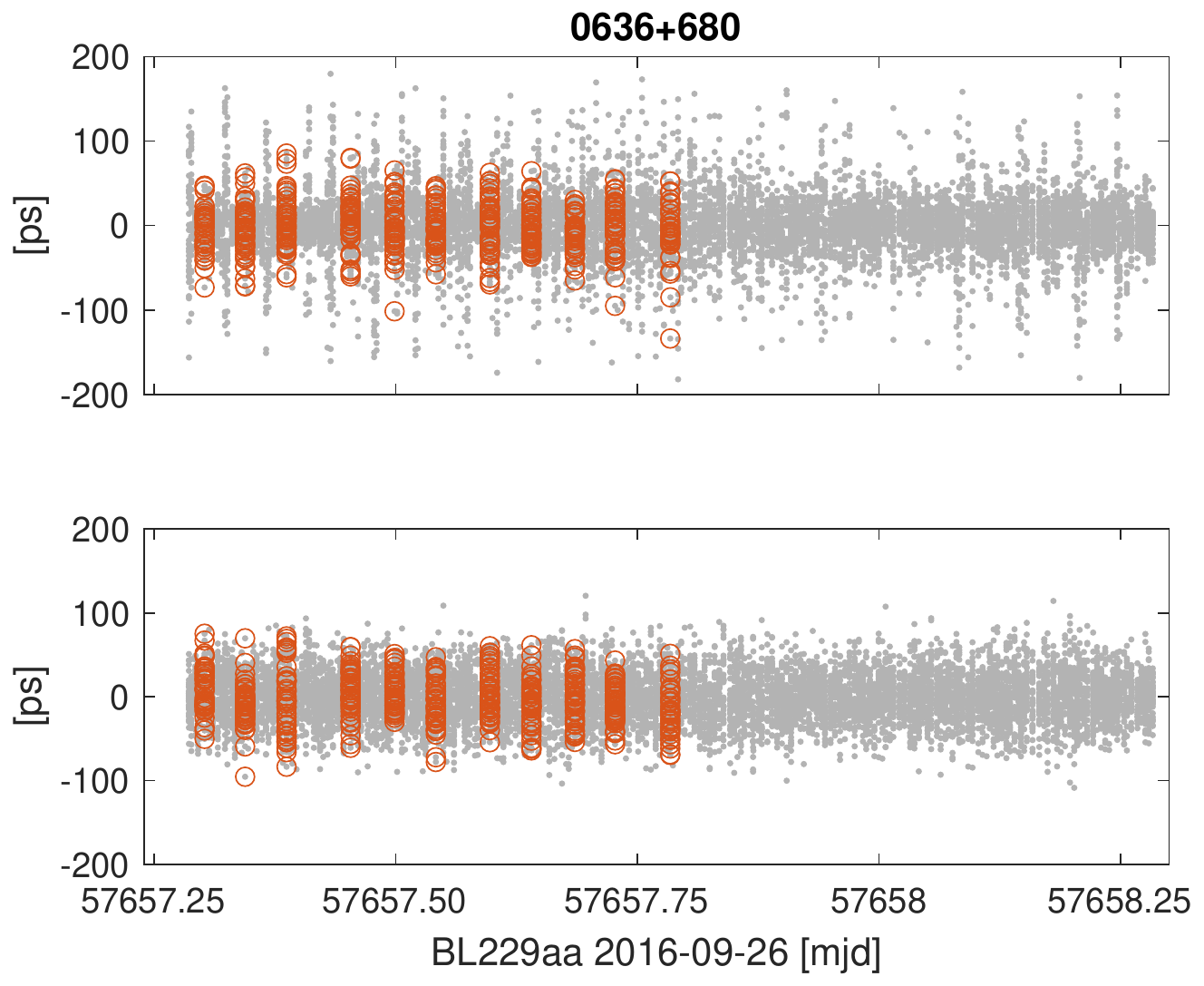}
    \includegraphics[width=0.48\textwidth]{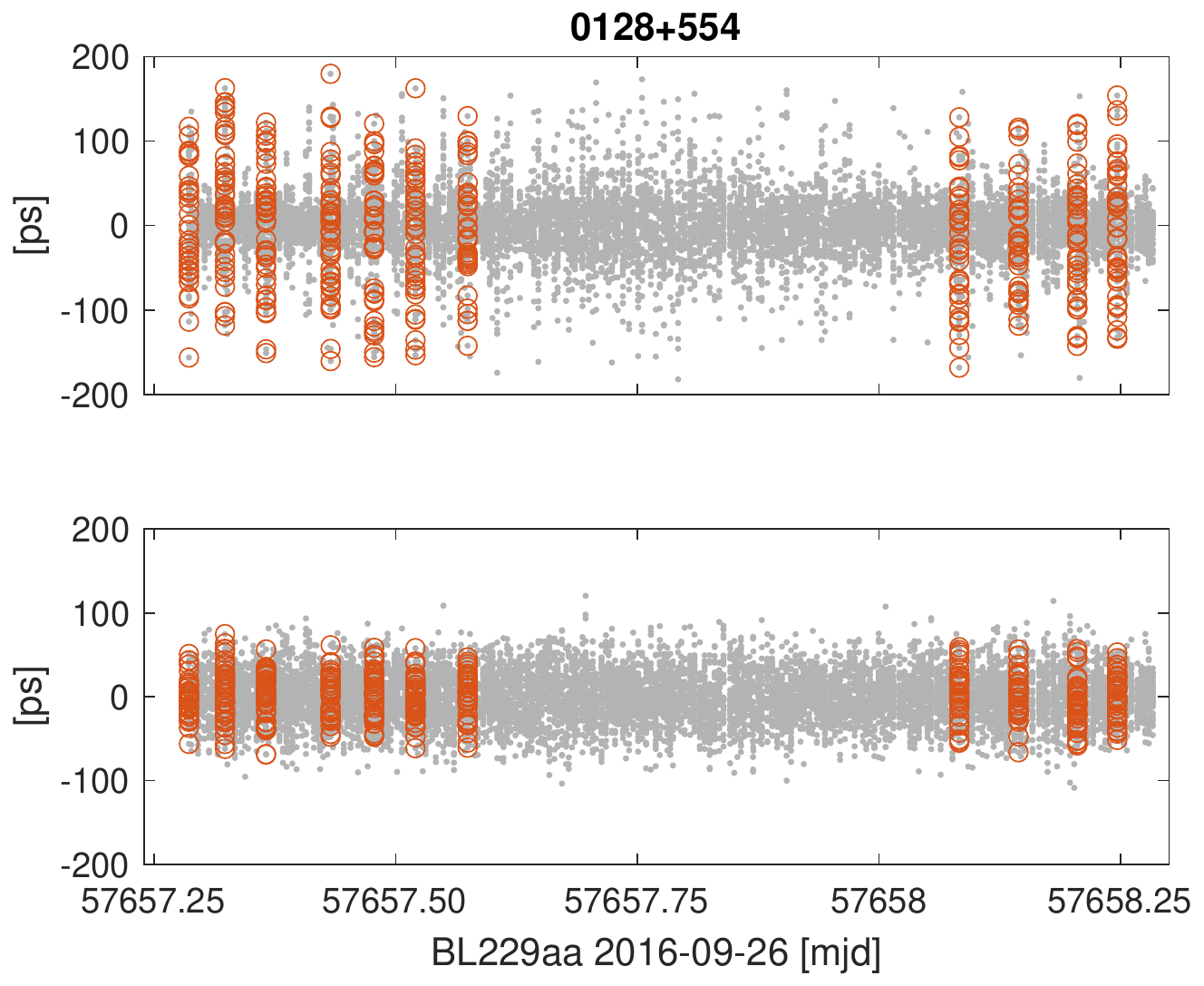}
    \caption{Post fit residuals in session bl229aa for real (upper plots) and simulated (lower plots) observations. Highlighted are sources 0636+680 with SI~2 on the left and 0128+554 with SI~4 on the right.}
\label{fig:residuals_sim}
\end{figure*}

The first indicator of astrometric source quality based on source structure
corrections was developed by \citet{Fey97}. They derived the so-called
structure index (SI) from VLBI source images as the median value of the
group structure delay ($\tau_{str}$) determined for pixels in
a $512 \times 512$ $uv$-grid for all baselines less than the diameter of the Earth.
Fig.~\ref{fig:souimage} shows images of the three most observed sources in
bl229aa (0636+680, 0210+515 and 0128+554) and in rv119 (2229+695, 0345+460,
0529+483) which are the first sessions of our datasets. We computed the SI for all sources
observed in bl229aa. For that calculation we used maps provided by MOJAVE
team\footnote{Available at \href{https://www.physics.purdue.edu/MOJAVE}{https://www.physics.purdue.edu/MOJAVE}}
and split the sources in the four SI groups according to the median value
of calculated structure delay corrections. Among 30~sources observed
in bl229aa, 3~sources have SI~1, 14~sources SI~2, 8~sources SI~3, and
5~sources have the highest SI~4. Fig.~\ref{fig:residuals_sim} shows post
fit residuals in session bl229aa for real observations $v_{real}$
(lower plots) and for simulated observations $v_{sim}$ (upper plots).
As an example, we highlight the most observed source 0636+680 in this
session which has structure index~2 and 0128+554 with structure index~4.
The comparison shows that the scatter of delay residuals for 0636+680 is similar
to real and simulated observations and the rms reaches 29.6~ps and 29.3~ps, respectively.
The rms of delay residuals from source 0128+554 with extended structure is 2.5~times
larger from real observations compared to the simulated ones, i.e. 72.3~ps and 27.4~ps, respectively. We computed the rms of delay residuals for every source in the bl229aa experiment and built the difference between the rms from real and simulated observations.
The median value of the rms difference was calculated as
\begin{equation}
   \Delta {\rm rms_{med}} = {\rm med (rms}(v_{real}) \; - \; {\rm rms}(v_{sim}))
\end{equation}
over each source group with respect to the structure index. The obtained median values are summarized in Table~\ref{tab:sim_residuals}.
We see that the difference between simulated and real delay residuals
is raising with an increasing source structure index since the structure
group delay is not modeled in the simulated observations. With the SI~1
taken as reference, the rms of the delay residuals increases by about 36~ps
for sources with SI~4. The lower wrms of postfit residuals from processing
real observations compared to the simulated ones for sources with low structure
indices (SI~1 and SI~2) that is manifested by the negative $\Delta {\rm rms_{med}}$
is due to the fact that the excessive noise due to source structure
is less than the random Gaussian noise with the rms of 20~ps that had been
added to simulated path delay in our simulation.

  We see that the source structure contribution increases the rms
of the postfit residuals but such an increase even for a subset of
sources with strong radio jets picked up for an astronomical program
does not have a noticeable impact on baseline length repeatability.
Source structure causes not only random but also systematic errors, but their impact on the baseline length repeatability is insigniﬁcant.
We exercise a caution to extrapolate this result to source position
estimates. This requires a further investigation that is beyond the
scope of present work.

\begin{table}
   \caption{Median of rms differences $\Delta {\rm rms_{med}}$ between delay residuals from real and simulated observations in bl229aa. Sources are divided in four groups according to their structure index, i.e. according to their median group structure delay $\rm med (\tau_{str})$. $N_{\rm sou}$ stands for number of sources in the group.}
   \label{tab:sim_residuals}       
   \begin{tabular}{rrrr}
         \hline\noalign{\smallskip}
         SI & $\rm med (\tau_{str})$ [ps]  & $N_{\rm sou}$ & $\Delta {\rm rms_{med}}$ [ps] \\
         \noalign{\smallskip}\hline\noalign{\smallskip}
         1 & 0 -- 3& 3 & -11.3 \\
         2 & 3 --10&14 & -6.6 \\
         3 & 10 -- 30& 8 & 9.0 \\
         4 & 30 -- $\infty$ & 5 & 25.0 \\
         \noalign{\smallskip}\hline
    \end{tabular}
\end{table}

\section{Conclusions}

  We have processed 33 diurnal astronomical observing VLBI sessions at
15~GHz under program MOJAVE-5 and 28 diurnal VLBI geodetic dual-band observing
sessions at 2 and 8~GHz under programs RV and CN. Both observing sessions ran
at the same ten station VLBA network with baseline lengths in a range from 237 to
8612~km at approximately the same time interval 2016.7--2020.5.

  We found that while the wrms of post-fit residuals from MOJAVE-5 program
was lower than from RV\&CN, 16.37~ps versus 26.13~ps, important metrics of
the geodetic quality of solutions, such as baseline length repeatability
and wrms of the differences of the ERP with respect to the reference IERS C04
times series were a factor 1.3 to 1.8 worse. We investigated the origin
of these discrepancies. We have established that modeling the ionospheric path
delay using the GNSS TEC maps was adequate for processing 15~GHz data during
the Solar minimum, and the errors of these TEC maps did not affect baseline length
repeatability at a noticeable level. We investigated whether the source
structure can be a factor, since MOJAVE-5 targeted objects with strong radio
jets and we have not found evidence it affected baseline length repeatability.
Finally, we ran solutions with simulated right hand sides for both MOJAVE-5 and
RV\&CN programs. The stochastic model used for these simulations was almost the
same. We were able to reproduce discrepancies in baseline lengths and EOP
time series statistics.

We have established that the major factor that causes discrepancies in baseline length repeatability
is a more agile schedule of RV\&CN experiments that includes more scans at low
and high elevations at short time intervals 1--3 hours than astronomical
experiments. We showed that the correlation coefficients between the station vertical component and atmospheric zenith path delay increase with an increasing elevation cutoff angle. When we removed observations below 20--$25^\circ$ elevations in RV\&CN,
we got a similar repeatability as in MOJAVE-5 program.

  Although the use of single-band astronomical VLBI data from MOJAVE-5 program
for geodesy provided less accurate results than the use of VLBI data from
the dedicated geodesy RV\&CN campaign, the baseline length repeatability is
still below 1~ppb. This gives us a good estimate of the impact of remaining
systematic errors that are specific for MOJAVE-5. This very low level of systematic
errors confirms that MOJAVE-5 dataset is an excellent testbed for investigation
of the effect of source structure on astrometry and geodesy in full detail.

\begin{acknowledgements}

HK works within the Hertha Firnberg position T697-N29, funded by the Austrian
Science Fund (FWF). LP work was supported by the NASA Earth Surface and Interior
program, project 19-ESI19-0030. This research has made use of data from the MOJAVE
database that is maintained by the MOJAVE team~\citep{Lister18}.
The Very Long Baseline Array (VLBA) is operated by the National Radio Astronomy Observatory, which is
a facility of the National Science Foundation, and operated under cooperative
agreement by Associated Universities, Inc. The authors acknowledge use of
the VLBA under the US Naval Observatory's time allocation.
This work supports USNO's ongoing research into the celestial reference frame
and geodesy.
\end{acknowledgements}

\bibliographystyle{spbasic}      
\interlinepenalty=10000 
\bibliography{reference_krasna}   


\end{document}